\documentclass[conference]{IEEEtran}
\IEEEoverridecommandlockouts
\usepackage{cite}
\usepackage{amsmath,amssymb,amsfonts}
\usepackage{algorithmic}
\usepackage{graphicx}
\usepackage{textcomp}
\usepackage{xcolor}
\usepackage{hyperref}
\usepackage{setspace}
\usepackage[utf8]{inputenc}
\usepackage[T1]{fontenc}
\usepackage{graphicx}
\usepackage{lscape} 
\usepackage{multirow}
\usepackage{lineno}
\usepackage{multirow}
\usepackage{booktabs}
\usepackage{subcaption}
\usepackage{amssymb}

\def\BibTeX{{\rm B\kern-.05em{\sc i\kern-.025em b}\kern-.08em
    T\kern-.1667em\lower.7ex\hbox{E}\kern-.125emX}}
    
\makeatletter
\def\@IEEEpubidpullup{8\baselineskip}
\makeatother

\begin{document}

\title{Modeling the impact of external influence on green behaviour spreading in multilayer financial networks\thanks{
        This work was supported by the Polish National Science Centre, grant no. 2016/21/D/ST6/02408 (PB, JJ) and 2018/31/B/HS4/00570 (MZ, AS). We would like to acknowledge the support of prof. Thomas Lux (University of Kiel) in the understanding multilayer bank-companies network model and sharing the data on connections between banks, and banks and companies.
    }
}

\author{

\IEEEauthorblockN{1\textsuperscript{st} Magdalena Zioło}
\IEEEauthorblockA{\textit{Department of Economics, Finance and Management} \\
\textit{University of Szczecin}\\
Szczecin, Poland \\
0000-0003-4136-046X}
\and

\IEEEauthorblockN{2\textsuperscript{nd} Piotr Bródka}
\IEEEauthorblockA{\textit{Department of Artificial Intelligence}\\
\textit{Wrocław University of Science and Technology}\\
Wrocław, Poland \\
0000-0002-6474-0089}
\and

\IEEEauthorblockN{3\textsuperscript{rd} Anna Spoz}
\IEEEauthorblockA{\textit{Department of Finance and Accountancy} \\
\textit{The John Paul II Catholic University of Lublin}\\
Lublin, Poland \\
0000-0001-5071-0017}
\and

\IEEEauthorblockN{4\textsuperscript{th} Jarosław Jankowski}
\IEEEauthorblockA{\textit{Department of Information Systems Engineering} \\
\textit{West Pomeranian University of Technology}\\
Szczecin, Poland \\
0000-0002-3658-3039}
}

\maketitle

\begin{abstract}
Growing awareness of the impact of business activity on the environment increases the pressure for governing bodies to address this issue. One possibility is to encourage or force the market into green behaviours. However, it is often hard to predict how different actions affect the market. Thus, to help with that, in this paper, we have proposed the green behaviour spreading model in the bank-company multilayer network. This model allows assessing how various elements like the duration of external influence, targeted market segment, or intensity of action affect the outcome regarding market greening level. The model evaluation results indicate that governing bodies, depending on the market "openness" to green activities, can adjust the duration and intensity of the proposed action. The strength of the impact can be changed by the public or private authority with the use of obligatory or voluntary rules and the proportion of influenced banks. This research may be helpful in the process of creating the optimal setups and increasing the performance of greening policies implementation.
\end{abstract}


\section{Introduction}
Green behaviour is of interest to various groups of entities, and it can be considered from the macro and microeconomic perspectives~\cite{ec2020taxonomy}. Among other sectors, financial institutions create initiatives to promote pro-environmental banking practices\cite{ifc2018sustainable}. Central banks and regulators want to contribute to improving environmental and climate risk management and gather capital for projects related to the transition to a sustainable economy~\cite{ngfs2021guide}. It was identified that the banking sector plays an important role in promoting sustainable development~\cite{bouma2017sustainable}. Adopting sustainable practices by banking institutions contributes to achieving the Sustainable Development Goals~\cite{prakash2018consolidation}. Initially, the social aspect was more visible in the activities of banks, but relatively quickly, the environmental aspect also gained importance~\cite{bukhari2019green}. It has been noticed that banks can have a direct and indirect impact on the natural environment~\cite{meena2013green}. Green banking is part of the broader concept of sustainable banking and focuses on promoting environmentally friendly practices\textcolor{green}~\cite{sbp2015concept}.

In the presented study, we focus on scenarios when the adoption of green products within banks is initialised by external influence from regulators~\cite{hahn2006approaches}, stakeholders~\cite{hahn2006approaches}, or mandatory sustainability guidelines~\cite{kale2015effects}. After implementing green financial products by banks, affiliated companies are also influenced to adopt green behaviour. This assumption is supported by earlier studies showing that banks may persuade clients to implement pro-ecological actions within their investments. The investments in environmentally friendly technologies may be connected with a better offer from banks\cite{bowman2010role}.

The main goal was to study the impact of the duration and the intensity of the external influence on banks on the diffusion of green behaviour among companies. Additional parameters were also taken into account related to the fraction of banks under external influence and to the impact of the threshold on the diffusion dynamics. The generalised model was proposed with a set of parameters which allow for adjusting it to a specific market. Simulations were performed within the network structures proposed in~\cite{lux2016model}. Results show the relation between external influence, its duration, its intensity, and the role of threshold on individual companies greening level. 

The paper is organized as follows: the introduction in section 1; in section 2, the theoretical aspects are presented. In section 3, the model structure and description of methods are described, and section 4 is the summary.

\section{Literature review}

The development of financial markets and transactions on these markets makes the financial system increasingly complex, involving many actors, including governments, financial institutions, businesses and households~\cite{li2022agent}. The network theory was used to describe the complexity of the financial system and to understand the functioning of the network of relations between system actors, its dynamics, scope and degree of impact~\cite{kenett2015network}. Examples of its use can be seen in numerous scientific studies in which networks were used: 1/ to describe the system structure~\cite{aleta2019multilayer}, 2/ to measure systemic features, e.g. resistance to certain scenarios or the impact of a specific policy on the operation of the system~\cite{monasterolo2020climate}, 3/ to assess the impact of the insolvency of an entity or a specific group of entities in the system, depending on their importance and connectivity within the structure~\cite{duan2019financial,battiston2012debtrank}, 4/ to assess the impact of liquidity problems initiated at different nodes of the system at specific time~\cite{haldane2011systemic} or 5/ to analyze systemic risk spreading~\cite{poledna2021quantification}.
Research has confirmed that financial systems are highly interdependent, which means that as the systemic risk increases, the stability of the network weakens~\cite{gong2019financial}. Features of financial networks have a strong impact on the risk spreading and market stability by influencing the individual agents' performances~\cite{acemoglu2015systemic}.
The banking system is a key element of the financial system. Aldasoro and Alves~\cite{aldasoro2018multiplex}, based on the network structure of large European banks, showed that banks that are well connected or important in one network are usually well connected in other networks as well. Analyzing the model of interactions between banks, Thurner et al.~\cite{thurner2003risk} demonstrated that well-connected networks are more effective in reducing risk and leading to fewer defaults. While examining the interbank network, Hüser~\cite{huser2015too} pointed to two main channels of the domino effect in the banking system: the domino effect through changes in the value of bank assets and direct links between interbank liabilities between financial institutions. The first channel is related to the network of claims and liabilities between individual institutions~\cite{langfield2012mapping,kenett2015network} and the second to the dynamics of loss propagation through a complex network of direct counterparty exposures after initial default~\cite{caccioli2018network,martinez2019interconnectedness}. Using a network approach, Óskarsdóttir and Bravo~\cite{oskarsdottir2021multilayer} showed that the default risk of an entity is highest when it is connected to many defaulting nodes, but the size of the neighbourhood of the entity mitigates the risk, showing both default risk and financial stability spreading throughout the network.
Based on ABMs, Li and Liu~\cite{li2022agent} simulated the real-world behavioural mechanisms of banks, companies, households, and the government and  studied the construction of a multiagent, multilayer endogenous financial network model. They observed the power-law distribution in supply chain network, the business credit network, the equity investment network, the bond investment network, and the interbank network. The deposit network and the loan network show a tendency of larger degree distributions for large banks and smaller degree distributions for small banks. Grilli et al.~\cite{grilli2020business} pointed out that the emergence of a large financial lending institution could lead to an over-centralization of the financial system, which had a negative impact on credit supply and the aggregate result. They emphasized that the regulator should pay more attention to the architecture of the credit market and try to avoid network concentration by redirecting the credit system to small local banks. This would help to avoid the effects of the impasse and the credit crisis and facilitate the monitoring of borrowers' financial conditions. Battiston et al.~\cite{ battiston2007credit} drew attention to the importance of the credit network economy based on the analysis of the effects of local interaction between firms connected by production and credit links. They showed that delay in payments could activate the effect of bankruptcy propagation. Interactions between firms have been the subject of research by de Jeude et al.~\cite{jeude2019multilayer}. They showed that companies directly or indirectly (corporate control, the influence of the management board, links between various structures, supply networks) influence each other.

Network features affect the speed and scope of spreading phenomena, including green behaviour. In a complex financial network, network nodes are interconnected by two-way relations~\cite{gong2019financial}. It is extremely important from the point of view of the propagation of phenomena and the possibility of managing them. Diebold and Yılmaz,~\cite{diebold2014network}, and Wang et al.~\cite{wang2021multilayer} pointed out that modelling  transmission of information and interconnections in the financial system is extremely difficult, and mapping complex financial systems into a single network structure may result in the simplification or loss of information. 

Recently several models of green behaviour spreading were proposed, and the factors influencing individuals’ green behaviour were analyzed~\cite{li2020conformity}. Gao and Tian~\cite{gao2019effects} investigated the diffusion of innovative knowledge in complex networks as a kind of green economic behaviour that maximizes the benefits obtained in the knowledge diffusion process. The results show that knowledge is more easily spread across heterogeneous networks and nodes with higher degrees. Li et al.~\cite{li2019impacts} showed that unquestioningly acquiring and believing in external information by individuals does not favour the diffusion of green behaviour. Another study based on the multiplex networks and the Microscopic Markov Chain Approach (MMCA) analyzed how information diffusion influences green behaviour. The results show that little initial information can trigger a sharp adoption of green behaviour. In another study, the impact of negative diffusion of information (about green behaviour) on the spread of green behaviour was analyzed~\cite{li2018impact}. Based on MMCA and Monte Carlo (MC) simulations, it has been shown that a slight influence of the information layer could hinder the surge of green behaviour. Consequently, controlling the dissemination of negative information helps spread green behaviour.

The banking sector plays a significant role in supporting sustainable development~\cite{bouma2017sustainable} as well as in adapting environmentally friendly strategies, mitigating climate risk and supporting the implementation of pro-ecological projects by redirecting funds to climate-sensitive sectors~\cite{park2020transition}. Fahlenbrach and Jondeau~\cite{fahlenbrach2021greening} showed that the actions of central banks could play a significant role in the process of greening the country's financial system by implementing green policies and regulatory measures. 
There are several channels through which banks can influence the behaviour of economic entities. By acting as creditors, investors, heads of supply chain and advisors, banks may make their clients more sensitive to environmental issues or persuade them to take pro-ecological actions.

The study focuses on the impact of the duration and the intensity of the external influence on banks on the diffusion of green behaviour among companies, which is a new area of research.

\section{Model structure}

In the proposed model, we assume that each bank and company has been assigned a Greening Level ($GL$), which reflects bank/company level of green behaviour adoption~\cite{ghosh2020product}. The value of $GL$ ranges from 0 to 1, where 0 means that the node (bank/company) does not adopt green behaviour at all, 1 means that the node is fully green in all actions and products, and low values like 0.1 mean that the node has some engagement in green behaviour (some green products, green assembly line, uses energy from renewable resources etc.), e.g. 10\% of bank loans or offers are green, or the company covers 10\% of its energy needs from solar power. In the beginning, we assume that all banks and companies have starting greening level ($SGL$). It could be adjusted to any value (depending on the existing situation in the particular market). For our evaluation, we assumed zero for all nodes at the beginning. 

While the whole network has a total aggregated Greening Level equal to zero, no diffusion of green behaviour takes place. Next, $GL$ of banks can be increased due to external influence, and it creates the ability to diffuse green behaviour to companies through a network of bank-company connections. 

External influence can last some period of time demoted by External Influence Time ($EIT$). This influence can be associated with mandatory or voluntary sustainability guidelines from market regulators like Government, European Union, etc.~\cite{kale2015effects}. They start influencing the financial sector to change banks' support for green behaviour, such as limiting carbon dioxide production. Another example of influence can be societal pressure demanding a higher level of environmental responsibility. 

While pressure can last $EIT$ discrete time steps in each step, attempts are taken to increase the Greening Level of each bank/node. The external influence with probability $\alpha$ can increase their $GL$ by some predefined value $\delta$ (e.g. 0.05, 0.1). The $\alpha$ represents the strength of the external influence. For example, mandatory rules may have a stronger influence than voluntary. The impact from regulators can be higher than the impact of customers, incentivized actions (e.g. lover tax) will have a higher impact etc. Porter~\cite{porter1995toward} showed that proper environmental regulations could positively impact green technological innovation. 

Greening Level Increase $\delta$ represents the adoption rate and willingness to adapt to green behaviour. A study conducted by~\cite{alshebami2021evaluating} revealed that green banking practices positively impact the green image of Saudi banks. According to~\cite{maniu2021adoption}, the motivation for green practices among SMEs owners is legislation, environmental concern but also potential cost-savings, improved public image, employee retention, and attracting new customers.

Apart from local parameters $\alpha$ and $\delta$ representing the probability of success of an external influence attempt, another parameter, External Influence Probability ($EIP$), was added. It is based on the assumptions that only part of nodes is subject to external influence, for example, 25\% of nodes, which represents types of banks, types of companies or economic sectors. Some frameworks like Equator Principles are not obligatory, and as such, they carry the burdens like lack of transparency, the limited scope of public disclosures, insufficient accountability, liability, monitoring, and implementation~\cite{worsdorfer201510}. 

The green behaviour spreading model is based on the linear threshold model~\cite{kempe2003maximizing} where node adopts some opinion \textit{A} if some portion of its neighbours also has opinion \textit{A}. Threshold values represented by the $LT$ parameter are used for transmission between banks and companies as the main target of our study. ~\cite{shrivastava1994greening} and~\cite{smith1994greening}, in independent studies, highlighted the growing importance of environmental business education as a response to the growing expectations of society in terms of dealing with problems related to environmental protection, employee and public safety as well as health hazards resulting from corporate activities. ~\cite{kyriakopoulos2020investigating} revealed a positive effect of environmental education on environmental behaviour. 

As a result, $LT$ depends on environmental education, customer expectation, global attitudes and awareness related to green behaviour. Since diffusion also takes place from banks to banks, companies to companies and companies to banks, thresholds are used for those transitions as well. Based on the fact that $LT$ represents global trends, the same value for each diffusion direction is used. Parameters $\alpha$ and $\delta$ are used for each direction of diffusion in the same way as it was used for external influence on banks, but they have different interpretation. Probability and adoption rate $\alpha$ and $\delta$ among companies can be related to special incentives like cheap loans or green investments reimbursement (e.g. installing solar panels, going paperless etc.)~\cite{chandrasekar2005effect} or managerial altitudes same like for banks~\cite{donald2009green}.

According to the above assumptions, we calculate the neighbours' influence on a node $i$ in the following way:

\begin{equation}
  L_i~=~\sum_{j\in N_i}{d_j \cdot GL_j},
\label{eq:influence}
\end{equation}

where: 
\begin{itemize}
 \item $N_i$ is the neighbourhood of a node $i$;
 \item $d_j$ is node $j$ degree normalised to neighbourhood of a node $i$, i.e., if node $i$ has three neighbours $j$, $k$ and $l$ with degrees 5, 1 and 4 respectively the $d_j=0.5$.
 \item $GL_j$ is the $GL$ of a node $j$
\end{itemize}

Each spreading iteration follows the following steps:
\begin{enumerate}
 \item External influence is trying to raise the $GL$ of all banks selected by External Influence Probability ($EIP$) with probability $\alpha$. If it succeeds, the nodes increase its $GL$ by $\delta$. All the following steps are calculated based on $GLs$ reached during this step.
 \item For each bank $j$, we calculate its $L_j$ for its neighbours in the bank layer, and if $L_j$ is above the predefined threshold $LT$, we increase its $GL$ by $\delta$. 
 \item For each bank $j$, we calculate its $L_j$ for its neighbours in the companies layer, and if $L_j$ is above the predefined threshold $LT$, we increase its $GL$ by $\delta$. 
 \item For each company $j$, we calculate its $L_j$ for its neighbours in the company layer, and if $L_j$ is above the predefined threshold $LT$, we increase its BL by $\delta$. 
 \item For each company $j$, we calculate its $L_j$ for its neighbours in the bank layer, and if $L_j$ is above the predefined threshold $LT$, we increase its $GL$ by $\delta$. 
 \item We update $GL$ for all banks and companies affected by the spreading process and go to step 1.
\end{enumerate}

\section{Numerical simulation and results}

In the next step, the simulation environment based on the proposed model was implemented as an agent-based model. It delivers functionalities required to perform simulations within network structures treated as a set of agents with possible interactions between them. The parameters used to evaluate the proposed model create experimental space presented in tab.~\ref{tab:parameters}. The number of banks and companies is based on the network proposed by~\cite{lux2016model} with 250 banks and 10~000 companies. The ratio between them is close to the relationship between banks and firms in real systems (see sec.~\ref{subsec:multilayer}). 

\begin{table}[b]
\centering
\caption{Parameters of diffusion used in simulations}\label{tab:parameters}
\begin{tabular}{|c|p{3.3cm}|p{3.5cm}|}
\hline

 & Parameter & Values\\\hline
$SGL$ & Starting greening level & 0\\\hline
$\alpha$ & Influence strength & 0.05, 0.1 \\\hline
$\delta$ & Adoption rate / Greening level increase & 0.05, 0.1\\\hline
$EIP$ & External influence probability & 0.25, 0.5, 0.75, 1.0 \\\hline

$EIT$ & External influence time & 1, 2, ..., 15 \\\hline
$SS$ & Simulation steps & 100 \\\hline

$LT$ & Linear threshold & 0.05, 0.1, 0.15, 0.2, 0.25 \\\hline

\end{tabular}

\end{table}

A set of implemented parameters makes it possible to execute various simulation scenarios and analyse the relation between input and output (greening level within the network and number of influenced companies). 

\subsection{Model of bank--company multilayer network}\label{subsec:multilayer}

To model the spread of green behaviour in the banking sector and connected companies, the multilayer network of banks and companies contains connections between banks (bank layer), between companies (company layer), and between banks and companies (interlayer connections). 

The model of the bank layer was proposed by~\cite{montagna2015hubs}, and it is based on the balance sheet structure of banks. First, bank sizes (reflected by their assets) are assigned based on the power law distribution. Next, each bank's profile is calculated, i.e., its external assets, interbank loans, net worth, and interbank liabilities. Finally, the banks are connected with probability which depends on the two banks' profiles (details in~\cite{montagna2015hubs}).

The model was extended in~\cite{lux2016model} to include relationships between banks and companies. A key part of the~\cite{lux2016model} model is the balance sheet allocation at the bank level based on Pareto distribution. The model is based on the $N_b$ number of banks and $N_f$ number of firms (companies). Number of links in firm sector to bank sector is denoted by $\lambda_f$ and links in bank sector to firm sector by $\lambda_b = \lambda_f N_f / N_b$. The distribution of degrees is not uniform. Bank-firm linking probabilities are related to the balance sheets of banks and loan sizes of firms. Balance sheet sizes for banks are based on truncated Pareto distribution applied to assets $A_i$, $i=1,2,...,N_b$. Degrees for banks in bank-firm connections are distributed in proportion to balance sheets with expected degree $\lambda_i = \bar{\lambda_b}$ $A_i$. Structure of firm-bank connections is based on loan sizes with mean loan size following: $\bar{f_j}=\bar\theta A_i N_b / N_f$ where $\theta$ represents the average fraction of external assets within balance sheet. Firm size distribution is based on the loan size distribution, and it influences the number of links to each firm from banks. More details and network visualisations are presented in~\cite{lux2016model}.

Presented models contain connections between banks (intralayer connections in the bank layer based on~\cite{montagna2015hubs}) and the connections between the bank and corporate sectors (interlayer connections between nodes in the bank layer and the company layer based on~\cite{lux2016model}). However, it does not contain connections between companies (intralayer connections in the companies layer). Thus, we have extended the model to include links between companies. To generate them we have use Barabási–Albert model~\cite{albert2002statistical} with $m=3$. Additionally, the companies have not been added to the network in random order, but they were ordered based on their size (taken from the~\cite{lux2016model} model), with bigger companies being added earlier to the network. As a consequence bigger, the company is the more connections to other companies it should have. 

\subsection{Impact of simulation parameters on greening level/Impact of influence parameters on greening level}

\begin{figure*}[t]
\centering
\begin{subfigure}[b]{0.24\textwidth}
  \includegraphics[width=\textwidth]{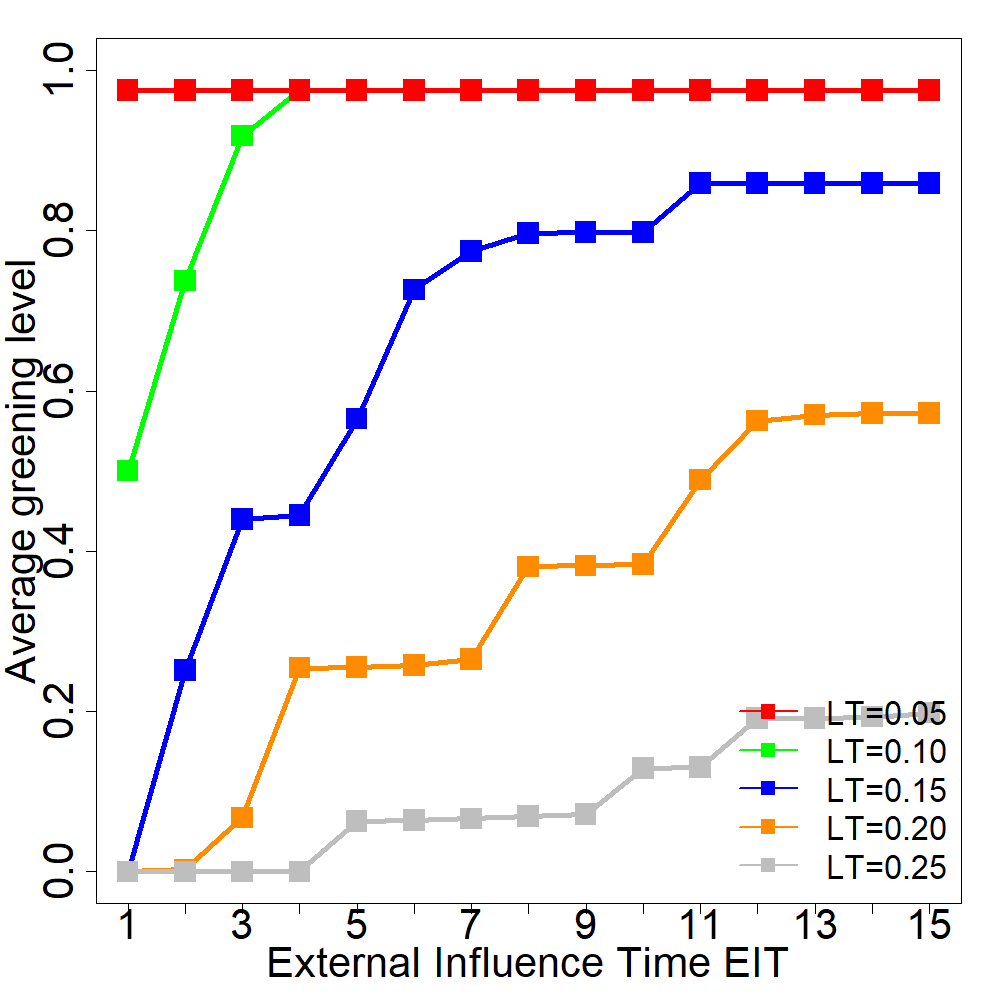}
  \caption{$LT$}
\end{subfigure}
\begin{subfigure}[b]{0.24\textwidth}
  \includegraphics[width=\textwidth]{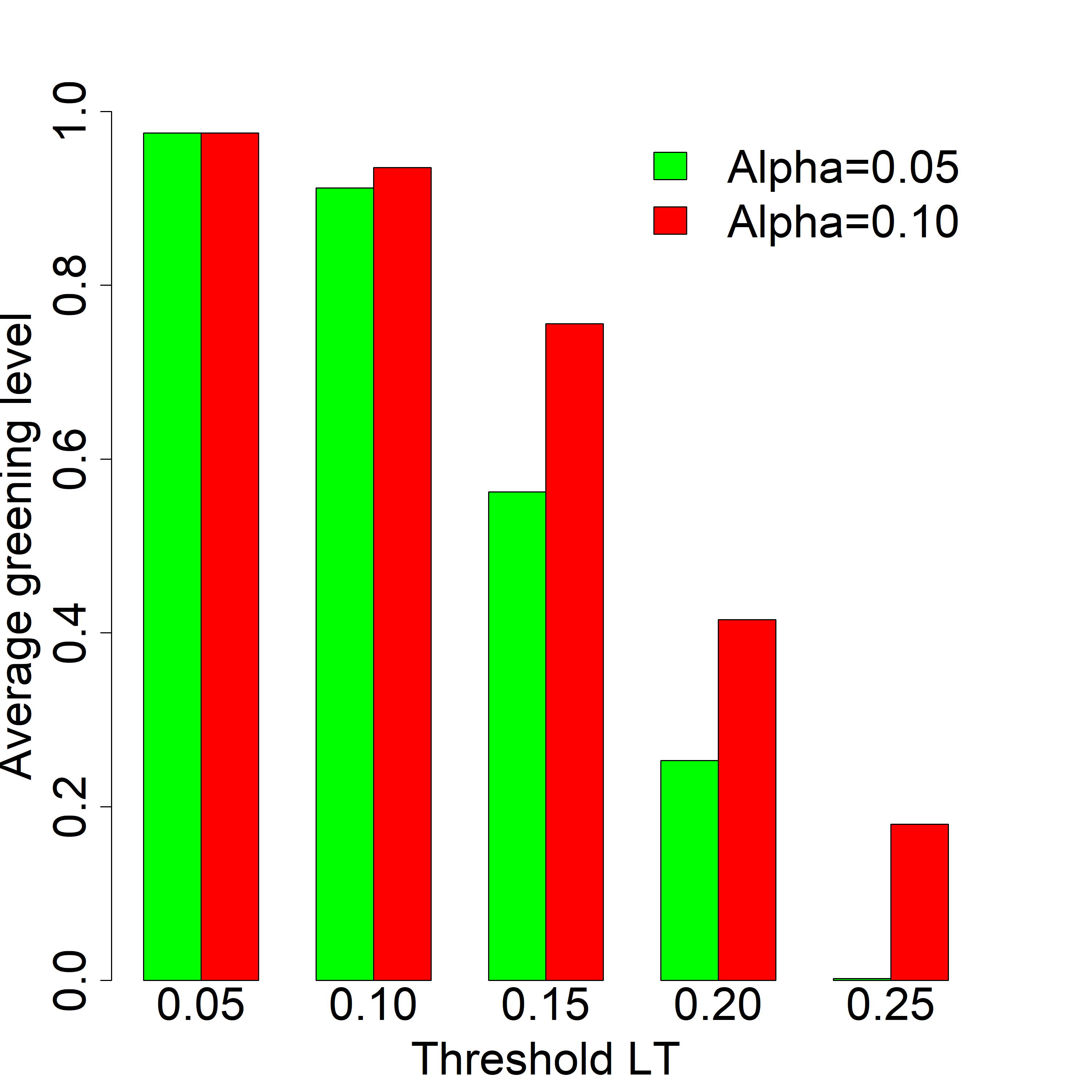}
  \caption{$\alpha$}
\end{subfigure}
\begin{subfigure}[b]{0.24\textwidth}
  \includegraphics[width=\textwidth]{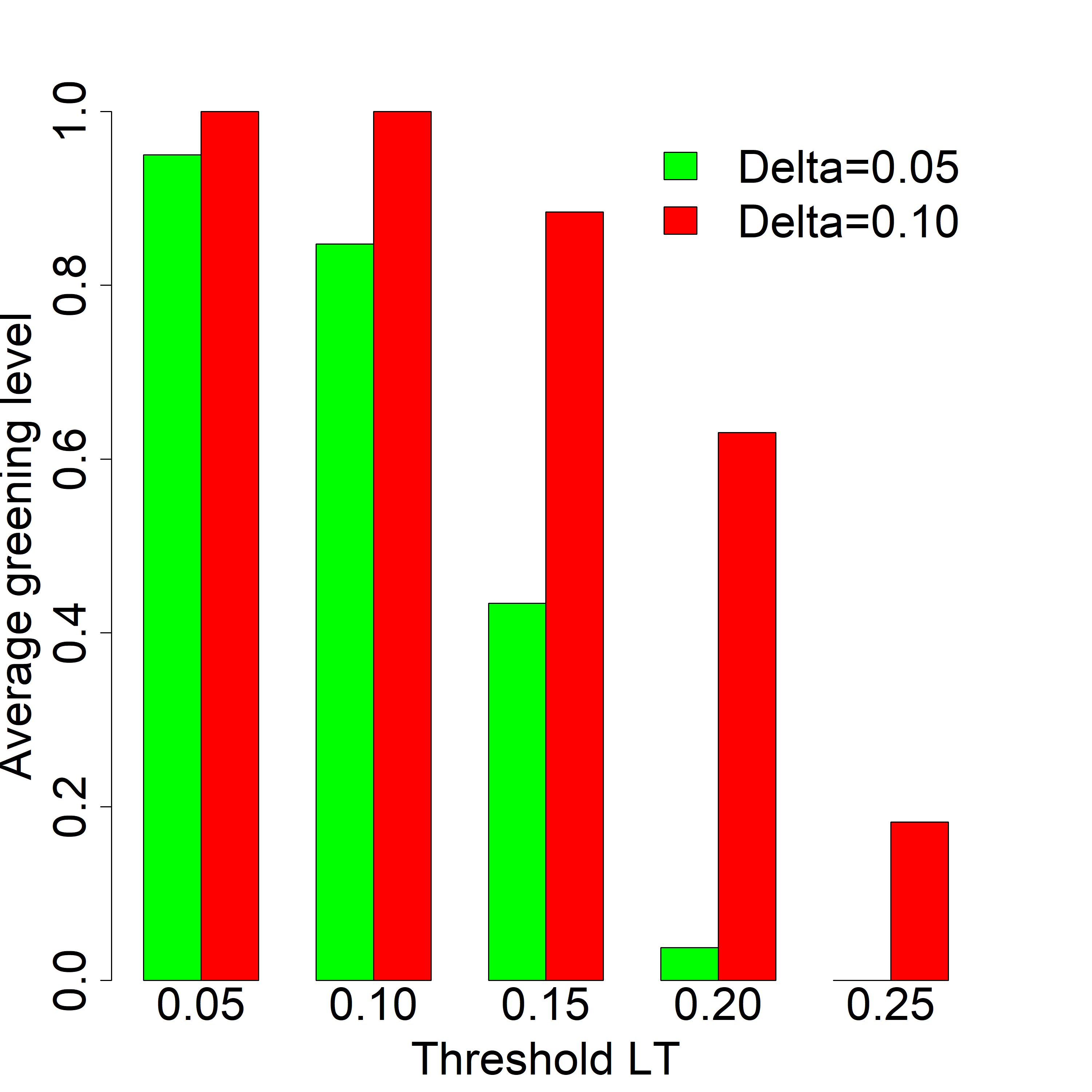}
  \caption{$\delta$}
\end{subfigure}
\begin{subfigure}[b]{0.24\textwidth}
  \includegraphics[width=\textwidth]{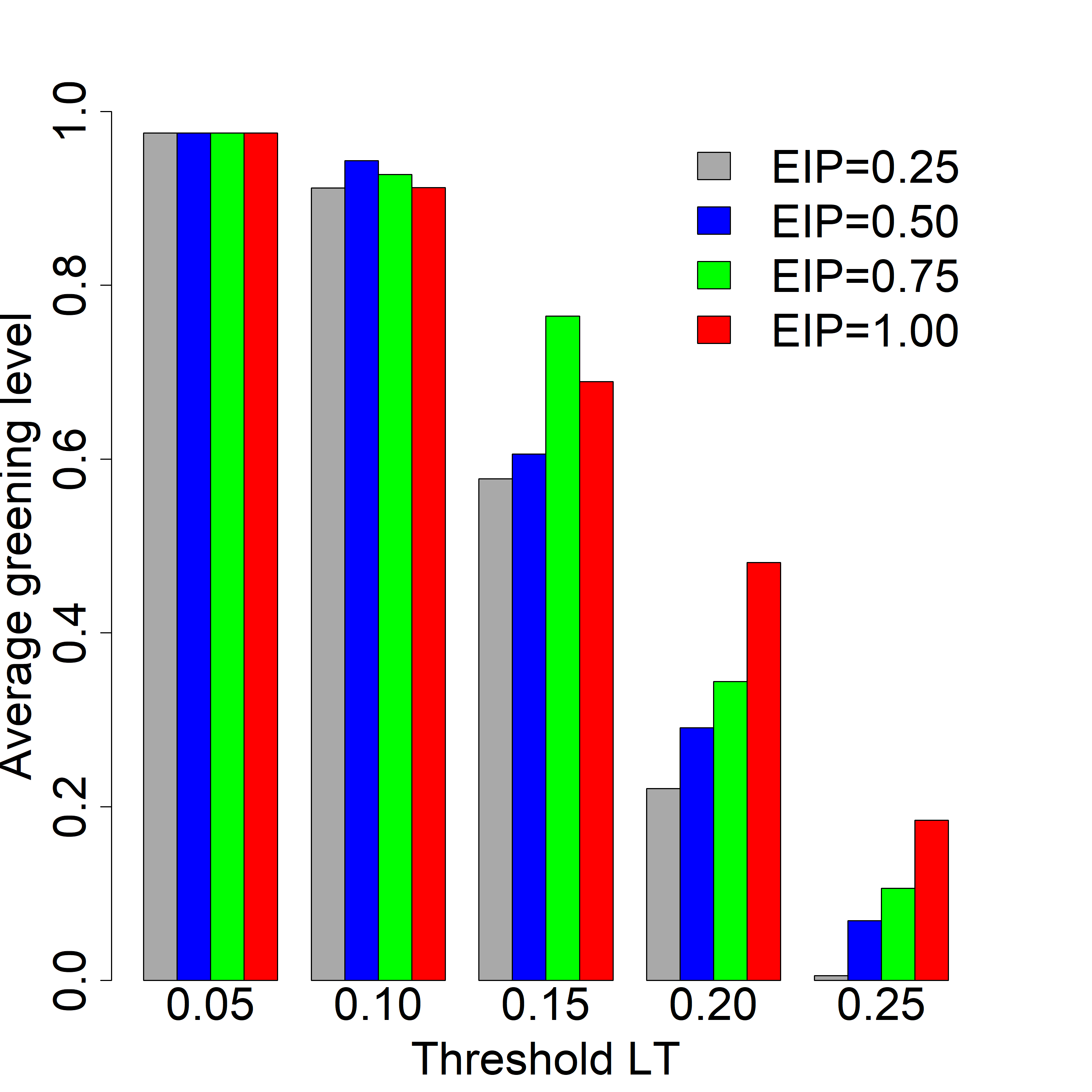}
  \caption{$EIP$}
\end{subfigure}

\caption{Average green value achieved by companies as a result of spread from banks, averaged in the last step from all simulation configurations for each threshold $LT$ and (a) External Influence Time, (b) $\alpha$, (c) $\delta$, (d) External Influence Probability.}
\label{fig:global_green}
\end{figure*}

The presented study was mainly focused on the effect of external influence on the banking sector in the first layer on the adoption of green technology by companies localised within the second layer. In the first stage average Greening Level ($GL$) value among companies was analysed. It was averaged at the and of simulation from the greening level achieved by companies. Results are presented in fig.~\ref{fig:global_green}. 

Impact of External Influence Time $EIT$ on greening level among companies for used values of $EIT$ (1--15) for all thresholds $LT$ is presented in fig.~\ref{fig:global_green}a. It shows the relation between reached average greening level values and time of external influence. For the lowest threshold, 0.05 maximal reached average greening level (1.0) is observed from the 2nd period of influence. The external influence was not required for longer than one period. While low $LT$ represents positive global altitudes, a high level of environmental education and low resistance on the node level, only a slight impact at the beginning was required to feed into network green approaches, and they flow quickly among all entities.  

If global attitudes and interest in green behaviour are lower and the threshold is increased twice to 0.1, we can still observe high diffusion dynamics. It was enough to influence the banking sector from $EIT$~=~1 to $EIT$~=~3. After that point and average $GL$ reached 0.92, slower growth is observed till the maximum $GL$ value of 0.98 in the 4th step. No additional external influence was required to achieve the highest greening level among companies. 

For higher thresholds representing lower global support and demand for green technologies, the impact of $EIT$ is clearly visible. Results in terms of average greening level are increased together with increased time of external influence. For threshold 0.15, the growth is observed till step 11, stabilising at the level of 0.86. The maximum value was not achieved, and external influence was not enough to overcome lower demand, education level and support for green behaviours. But still, the first stages of external influence resulted in high growth dynamics of the greening level. The longer influence did not improve results. More efforts could be made into building social awareness and impact on society than a direct influence on financial institutions. 

Worse global support for greening, lower demand and education level represented by $LT$ at the level of 0.20 requires higher external influence. It results in continuous improvement of greening levels. 

Highest used $LT$ (0.25) resulted in growth from $EIT$~=~4 based on three periods of stabilisation 5--9, 10--11, 12--15 at the level of 0.07, 0.13 and 0.19. Even though the difference between $LT$~=~0.20 and 0.25 is relatively low, the drop in performance is substantial. It shows how important the proper evaluation of the model results is for assessing market $LT$. 

Fig.~\ref{fig:global_green}b shows the impact of the $\alpha$ parameter on the average greening level within a network of companies. It represents the strength of external influence on banks and is related mainly to the type of applied regulatory rules. Analysis for each $LT$ representing global conditions and demand for green technologies shows that for the lowest resistance and linear threshold parameters like 0.05 and 0.10, differences between used $\alpha$ values 0.05 and 0.1 are very small. For $LT$~=~0.05 lowest $\alpha$ parameter 0.05 delivers average values of green behaviour at level 0.98, same as for $\alpha$~=~0.1. It shows that strong impact and obligatory rules are unnecessary if global support and demand for green technologies represented by the lowest $LT$ is strong.

For $LT$~=~0.10, results increase from 0.91 observed for $\alpha$~=~0.05 to 0.94. For setups with higher $LT$ with values 0.15, 0.20 and 0.25 increase of the $\alpha$ parameter was more justified as a substantial increase in average network greening level was observed. 

Another parameter, Greening Level Increase $\delta$, represents the green value delivered during external influence with a single contact. It represents the adoption rate and willingness to adopt green behaviour at the node level after influence attempts from another node. The simulation showed the role of the intensity of impact based on low level $\delta$~=~0.05 and the effect of is strengthening twice to 0.1. Fig.~\ref{fig:global_green}c shows impact of $\delta$ values for each $LT$ value. We need to remember that increasing $\delta$ to 0.1 from baseline 0.05 creates additional costs for a potential campaign; thus, it is important to be able to assess the effect of this change. For a setup with the lowest threshold of 0.05 (it is easy to increase the greening level), the impact of changing $\delta$ was very low and increased from 0.95 to 1.00. A similar situation we have for $LT$~=~0.1, the increase was from 0.85 to 1.00. Substantial effect of the $\delta$ increase, started from $LT$~=~0.15 with increase of 204\%, and 1659\% for $LT$=0.20. The highest gain was for $LT$~=~0.25, with the increase of average green from just 0.000083 to 0.18. 

Apart from $\alpha$ and $\delta$ parameters representing the strength of communication, from the perspective of external influence, we have yet another important variable, namely, the size of the target group. Several strategies can be selected to influence the market. We can either target all banks or only a specific group of them. To represent such a situation, we used the External Influence Probability parameter (EIP) to initially select nodes for further influence attempts. $EIP$~=~1.00 represents the biggest coverage, and all banks are selected for influence. It usually has the highest costs since all institutions are influenced. Lower values 0.75, 0.50 and 0.25 represent probabilities of selecting specific groups of banks for external influence (in the case of our simulations, it reflects 75\%, 50\% and 25\% of the market). For example, in real campaigns, corporate banks can be selected or investment banks etc. Results for each used $EIP$ are presented in fig.~\ref{fig:global_green}d. For $LT$~=~0.05, no increase of average green was observed after targeting all banks instead of only 25\% of them. Results for $LT$~=~0.10 show small differences however the differences are not significant, and we cannot draw any conclusion whether it is worth increasing $EIP$ or not. For $LT$=0.15 changes of $EIP$ from 0.25 to 1.00 resulted increase from 0.22 to 0.48. Changes of $EIP$ from 0.25 to 0.50 and 0.50 to 0.75 showed increase from 0.22 to 0.29 and from 0.29 to 0.34. A further increase from 0.75 to 1.00 did not show a performance increase. Higher differences between all External Influence Probabilities are visible for thresholds 0.20 and 0.25. In those cases, increasing $EIP$ results in an increase in the average greening level.
For $LT$=0.20 changes of $EIP$ from 0.25 to 0.5 resulted increase from 0.22 to 0.29, changes from 0.50 to 0.75 showed increase from 0.29 to 0.34. Changes from 0.75 to 1.00 delivered increased from 0.34 to 0.48. The highest threshold of 0.25 resulted in even bigger differences with a 32.52 times increase in green average for $EIP$ changed from 0.25 to 1.00. Changes of $EIP$ from 0.25 to 0.50, 0.50 to 0.75, 0.75 to 1.00 showed increase equal to 12.48, 1.54 and 1.74 times.

There are several conclusions for the financial system from the conducted research. Above all, it is important to activate top-down involvement in raising awareness of green banking and its importance in achieving the goals of climate policy, including the transformation of enterprises towards sustainability. Regulators are aware of this fact; therefore, the number of legal regulations in this area is increasing every year, although their nature and scope vary (they can be of international, national or local scope). In developed countries, they mainly focus on the disclosure of information, while in developing countries, more detailed and strict regulations are implemented~\cite{park2020transition}. In countries with high environmental awareness, regulations may be voluntary, unlike in countries with low environmental awareness, where it is important to apply regulatory obligation to the entire collectivity. 

The greening drivers in an ecologically aware environment may be of a one-time and short-term nature, while in cases where the green culture is still being built, the tools for greening should be more personalized and strengthened over time. By creating an attractive green product offer, banks can create a supportive environment for enterprises to transform towards sustainability. Preferential loans for green investments and green credits should be related to the state's environmental policy, facilitating the achievement of these policy goals and strengthening its effects. In the process of building environmental awareness, initiatives such as the development of the EU taxonomy, certification and labelling, or the use of ESG ratings are extremely helpful. These initiatives enable the identification of the entity in the context of sustainability and facilitate investment decisions. They contribute to determining the cost of obtaining foreign capital, and in extreme cases, they can make it difficult or impossible to obtain the capital.

\subsection{Impact of diffusion process parameters on performance in simulation steps/Impact of diffusion process parameters on greening performance}

\begin{figure*}[t]
\begin{subfigure}[b]{0.31\textwidth}
  \includegraphics[width=\textwidth]{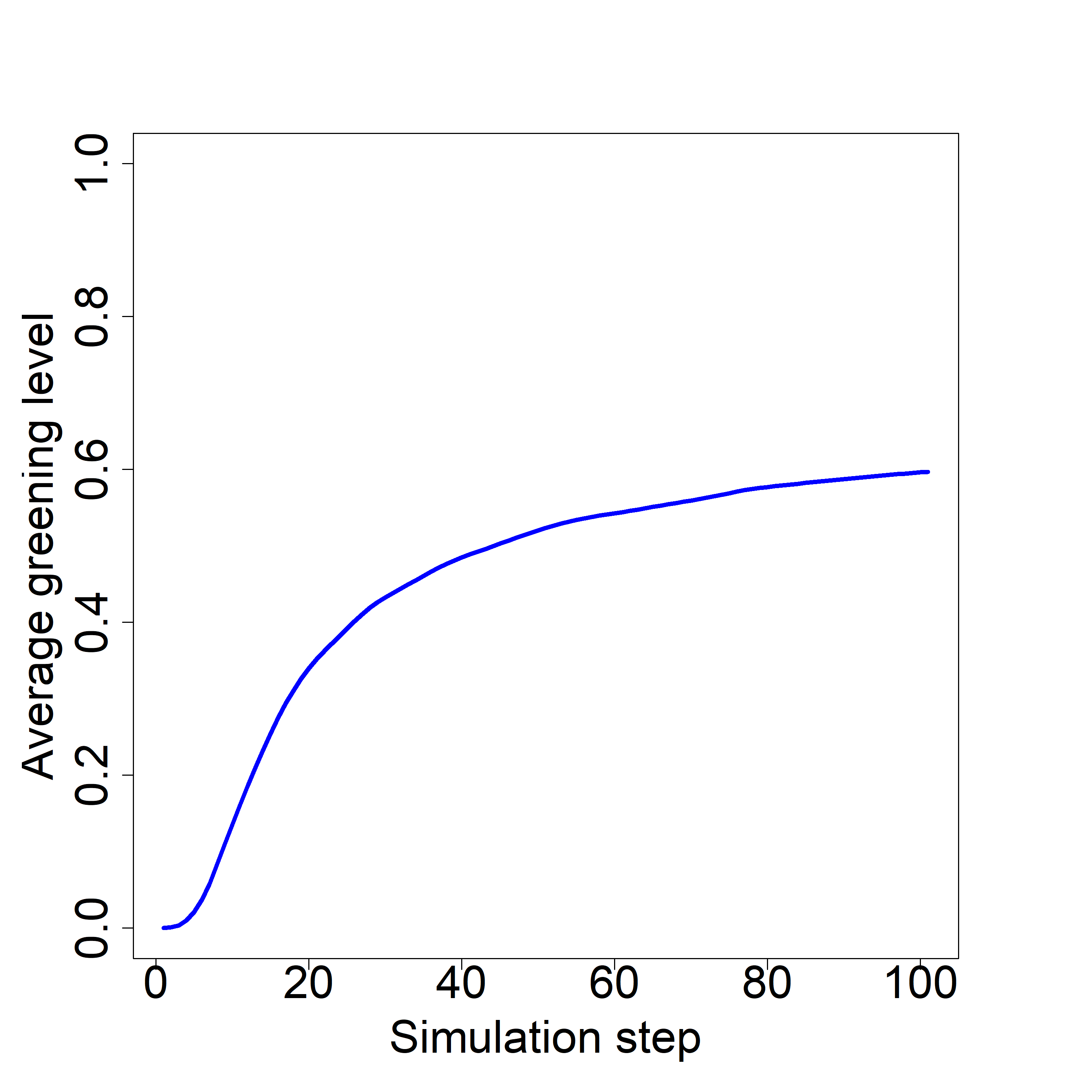}
  \caption{All simulation configurations}
\end{subfigure}
\begin{subfigure}[b]{0.31\textwidth}
  \includegraphics[width=\textwidth]{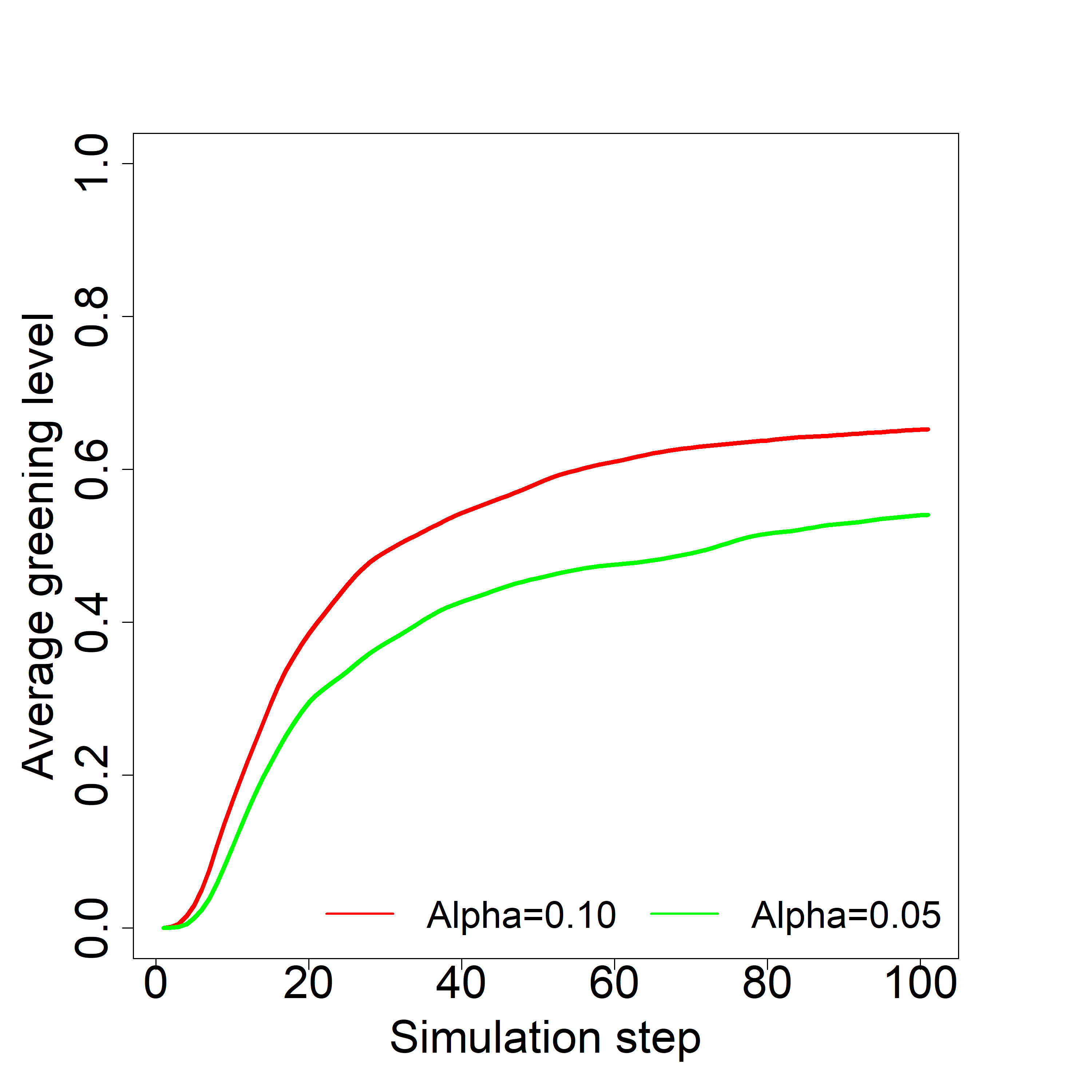}
  \caption{$\alpha$}
\end{subfigure}
\begin{subfigure}[b]{0.31\textwidth}
  \includegraphics[width=\textwidth]{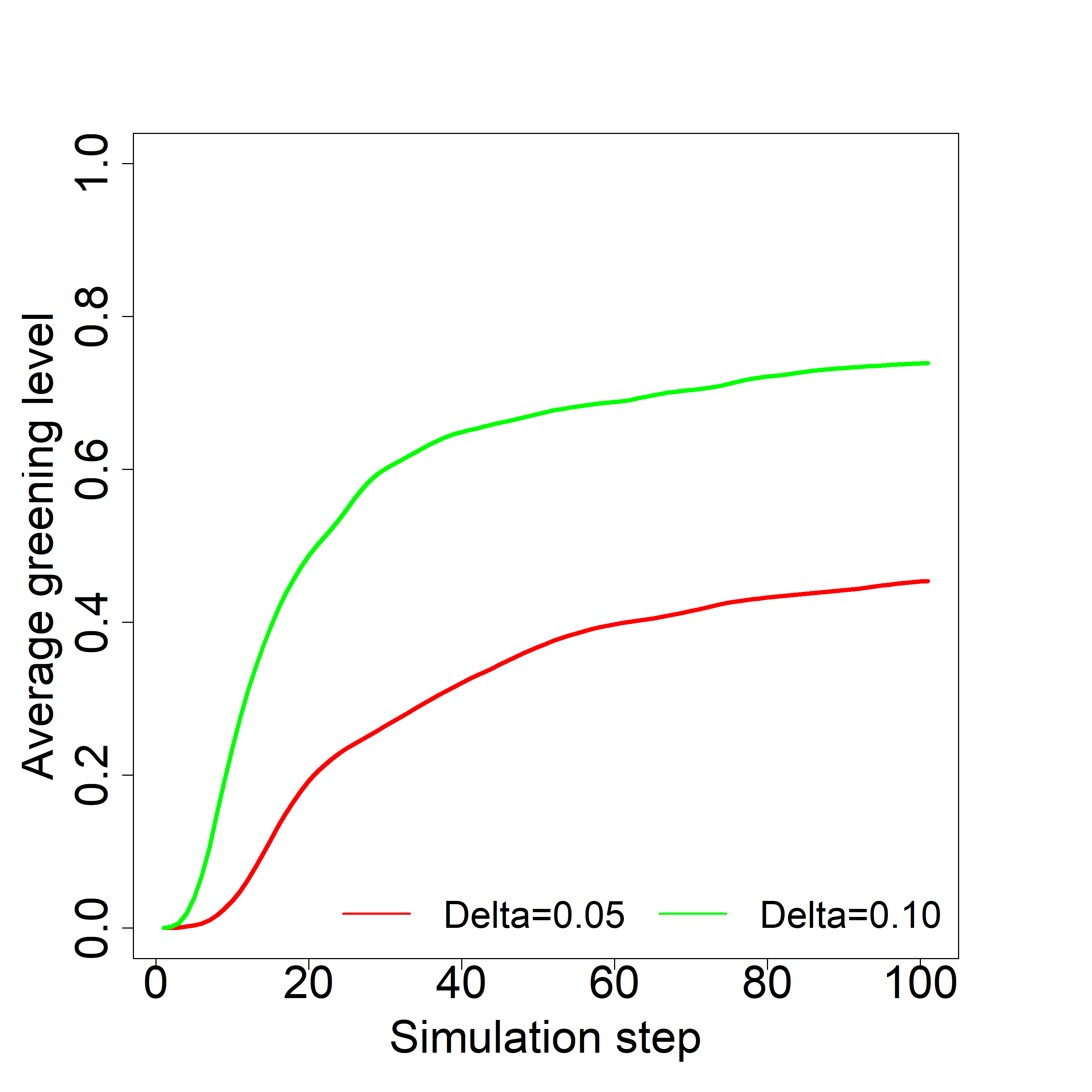}
  \caption{$\delta$}
\end{subfigure}
\begin{subfigure}[b]{0.31\textwidth}
  \includegraphics[width=\textwidth]{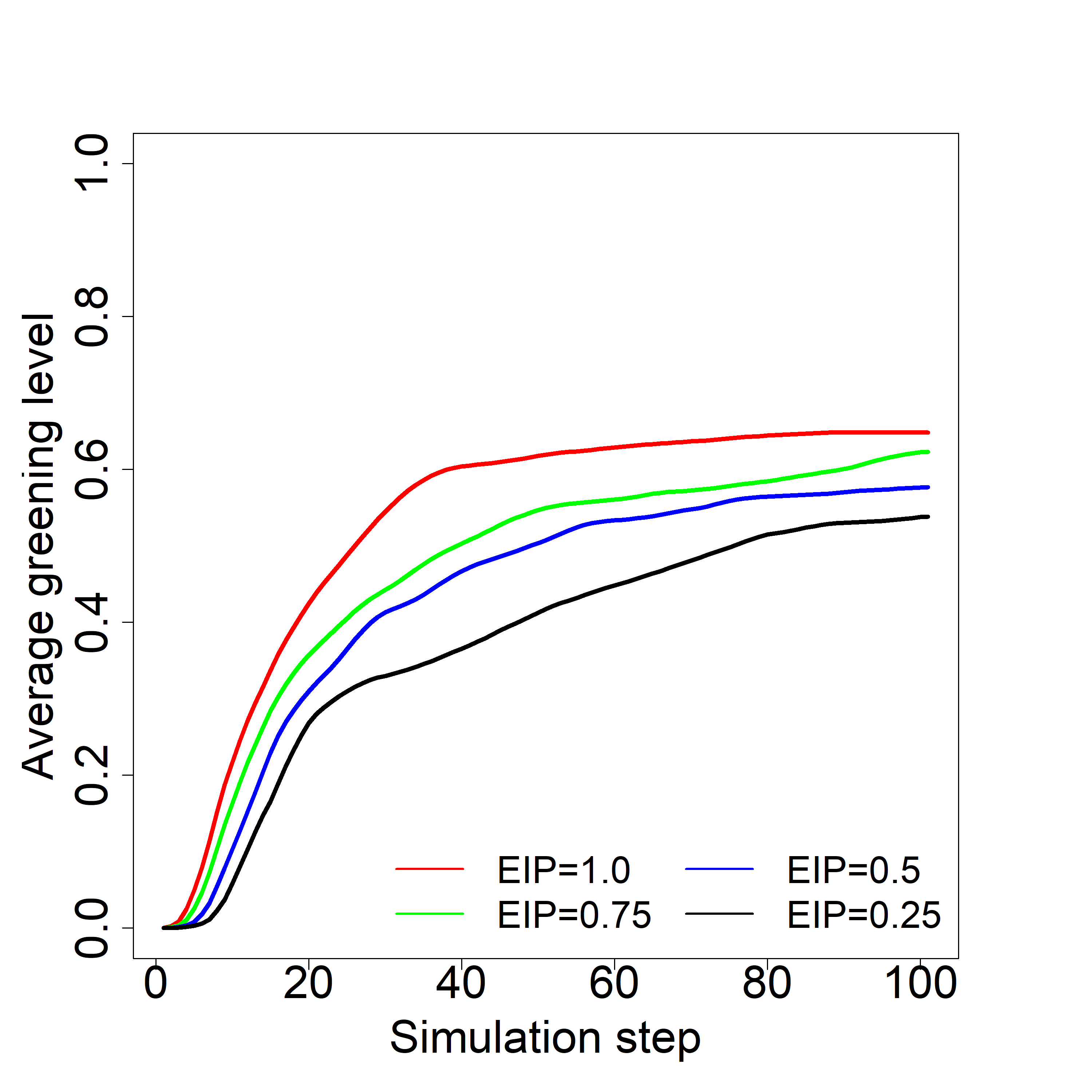}
  \caption{$EIP$}
\end{subfigure}
\begin{subfigure}[b]{0.31\textwidth}
  \includegraphics[width=\textwidth]{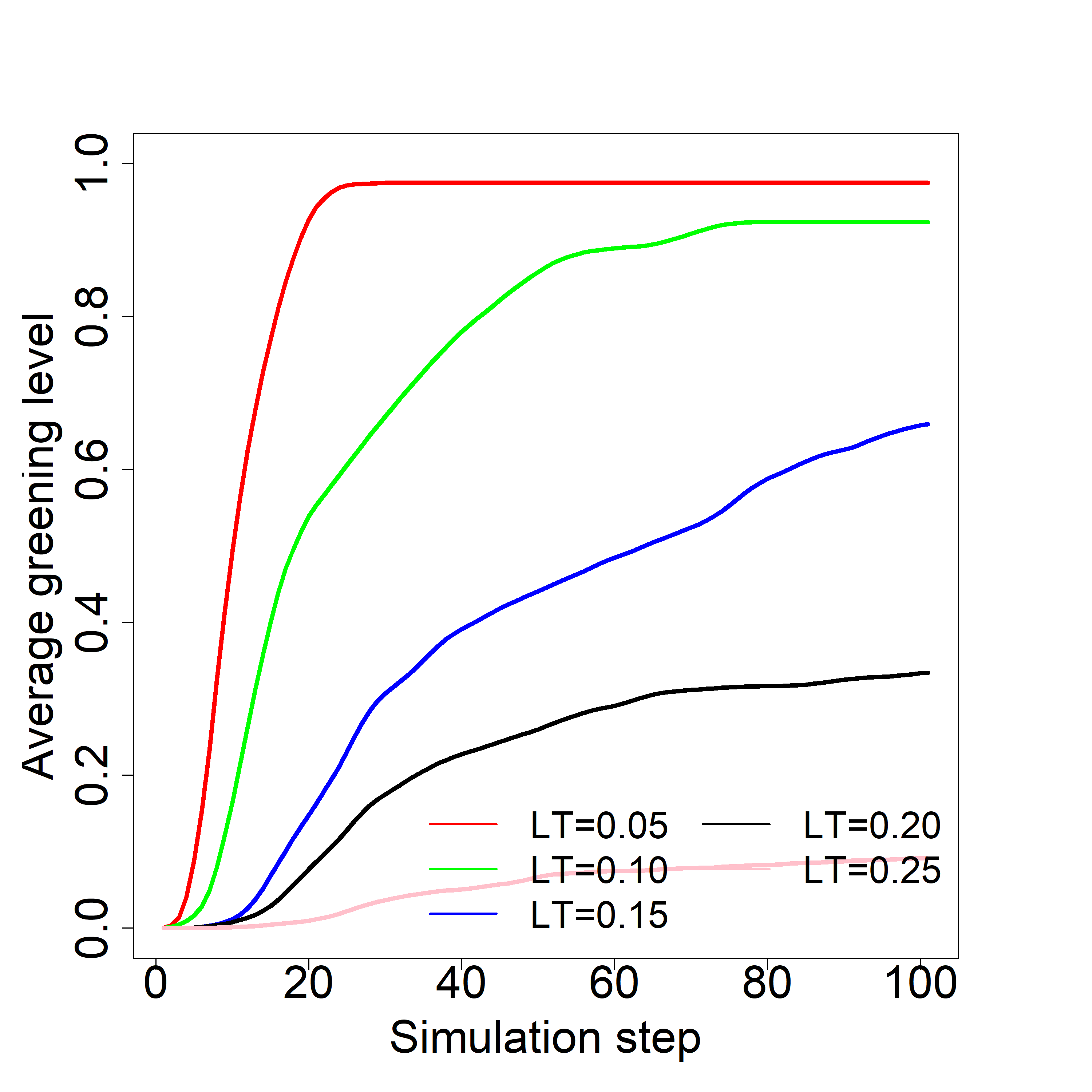}
  \caption{$LT$}
\end{subfigure}
\begin{subfigure}[b]{0.31\textwidth}
  \includegraphics[width=\textwidth]{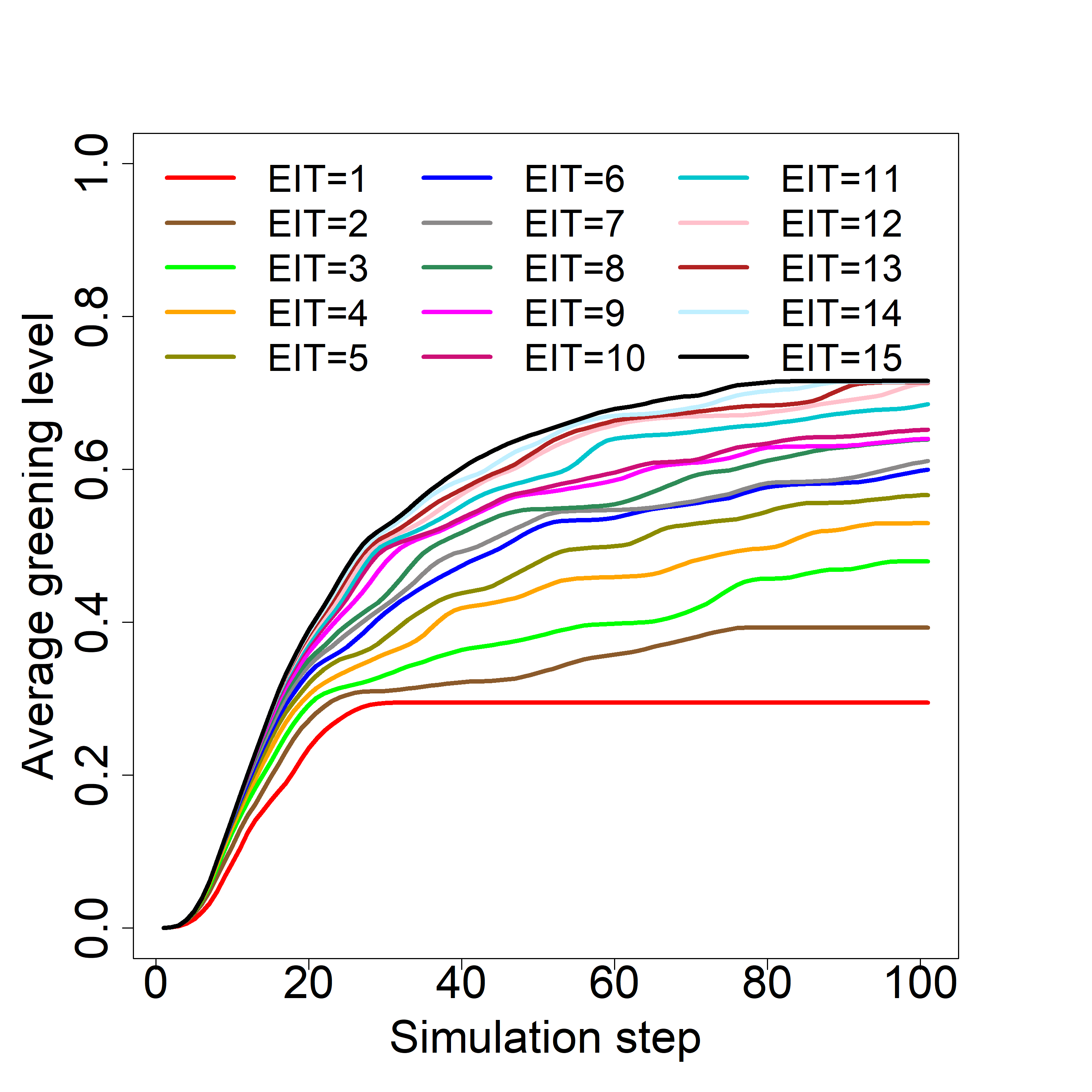}
  \caption{$EIT$}
\end{subfigure}
\caption{Average green value for companies in each simulation step as result of spread from banks for (\textbf{a}) all simulations, (\textbf{b}) $\alpha$, (\textbf{c}) $\delta$, (\textbf{d}) External Influence Probability $EIP$, (\textbf{e}) threshold $LT$, (\textbf{f}) External Influence Time $EIT$.}
\label{fig:steps_global}
\end{figure*}

 While analysis of the average greening level within a network at the end of the simulations shows the total impact of used parameters on the whole process, another dimension of analysis shows the process of spread over time for all steps of simulations. Overall results for whole parameters for spread to companies are shown in fig.~\ref{fig:steps_global}. The total average value of greening level is presented in fig.~\ref{fig:steps_global}(a) shows that the highest spread dynamics were observed till step 13 with more than 10\% increase in average greening level in each simulation step. Between step 13 and increase was above 5\% in each step and below 1\% starting from the 38th step. 

The diffusion dynamics were influenced by the $\alpha$ parameter with used 0.05 and 0.1 values. Results are presented in fig.~\ref{fig:steps_global}(b). Till step 5, differences in each step are at least two times higher for $\alpha$~=~0.1. Then till step 33, results for $\alpha$~=~0.1 are at least 1.3 times higher than for $\alpha$~=~0.05. From the 34th step, the difference drops from 1.29 to 1.21 in the last step. 

Results for Greening Level Increase $\delta$ show a growing difference from the beginning of the process with results visible in fig.~\ref{fig:steps_global}(c). The highest difference was observed in step 5 with 11.67 times higher average value of Greening Level for $\delta$~=~0.1. Till step 10 difference is at least 5 times higher, and it drops to 2.0 in the 40th step and to 1.63 in the last step. 

Impact of $EIP$ on the average green is presented in fig.~\ref{fig:steps_global}(d). Differences in results are higher at the beginning of the process. The highest difference between $EIP$~=~1 and $EIP$~=~0.75 was observed in step 2, with 2.37 times higher average green for $EIP$~=~1. The highest difference between $EIP$~=~0.75 and $EIP$~=~0.50 was observed in step 4, with a 3.26 times higher average green for $EIP$~=~0.75. The highest difference between $EIP$~=~0.50 and $EIP$~=~0.25 was observed in step 5, with 3.07 times higher average green for $EIP$~=~0.5.

Process dynamics was dependent on thresholds (fig.~\ref{fig:steps_global}(e)). The highest performance is observed for threshold 0.05, where in the 31st step, saturation at the level 0.98 is reached. Threshold 0.1 achieves a maximal value of 0.92 in step 72. For thresholds 0.15, 0.20 and 0.25, maximal levels are reached in steps 98 (with the average value of green 0.66), 90 (with the average value of green 0.33) and step 84 (with the average value of green 0.09). 

Impact of External Influence Time $EIT$ for all scenarios with 1 up to 15 steps of influence in each simulation step is shown in fig.~\ref{fig:steps_global}(f). An increase of $EIT$ shows the biggest differences in average green between low values of $EIT$ from 1 to 6. One step ($EIP$~=~1) allowed us to achieve a green value at the level of 0.30 in the 31st step of the simulation, and the spreading stopped. Increasing external influence time to 2 allowed to reach in 72nd step greening level at 0.39, 1.33 times higher than for $EIT$~=~1. A lower difference is observed between $EIT$~=~2 and $EIT$~=~3 with 1.22 times better avenge greening level. Further increase of $EIT$ delivers smaller changes at 1.1 and 1.07 after comparing $EIT$~=~4 vs $EIT$~=~3 and $EIT$~=~5 vs $EIT$~=~4.

Further increasing $EIT$ resulted in 1\%-5\% increase when $EIT$(i) was compared with $EIT$(i-1). 

While $EIT$ is associated with costs of influence, incentives or tax preferences, longer influence increases costs, and results showed that increasing the campaign duration two times (e.g. from 5 to 10) will not double the average greening level two times. 

\begin{figure*}[!ht]
\begin{subfigure}[b]{0.31\textwidth}
  \includegraphics[width=\textwidth]{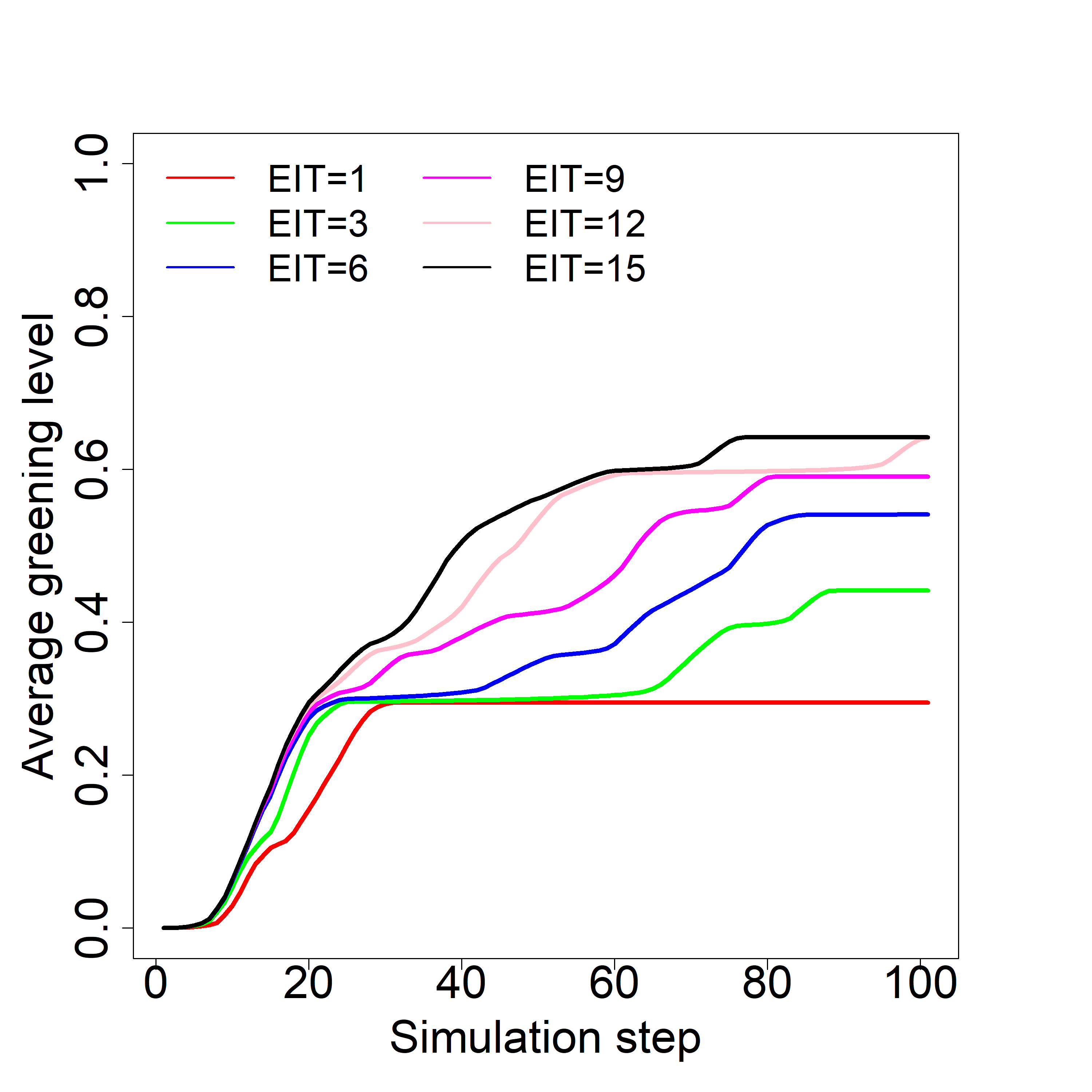}
  \caption{EIP=0.25}

\end{subfigure}
\begin{subfigure}[b]{0.31\textwidth}
  \includegraphics[width=\textwidth]{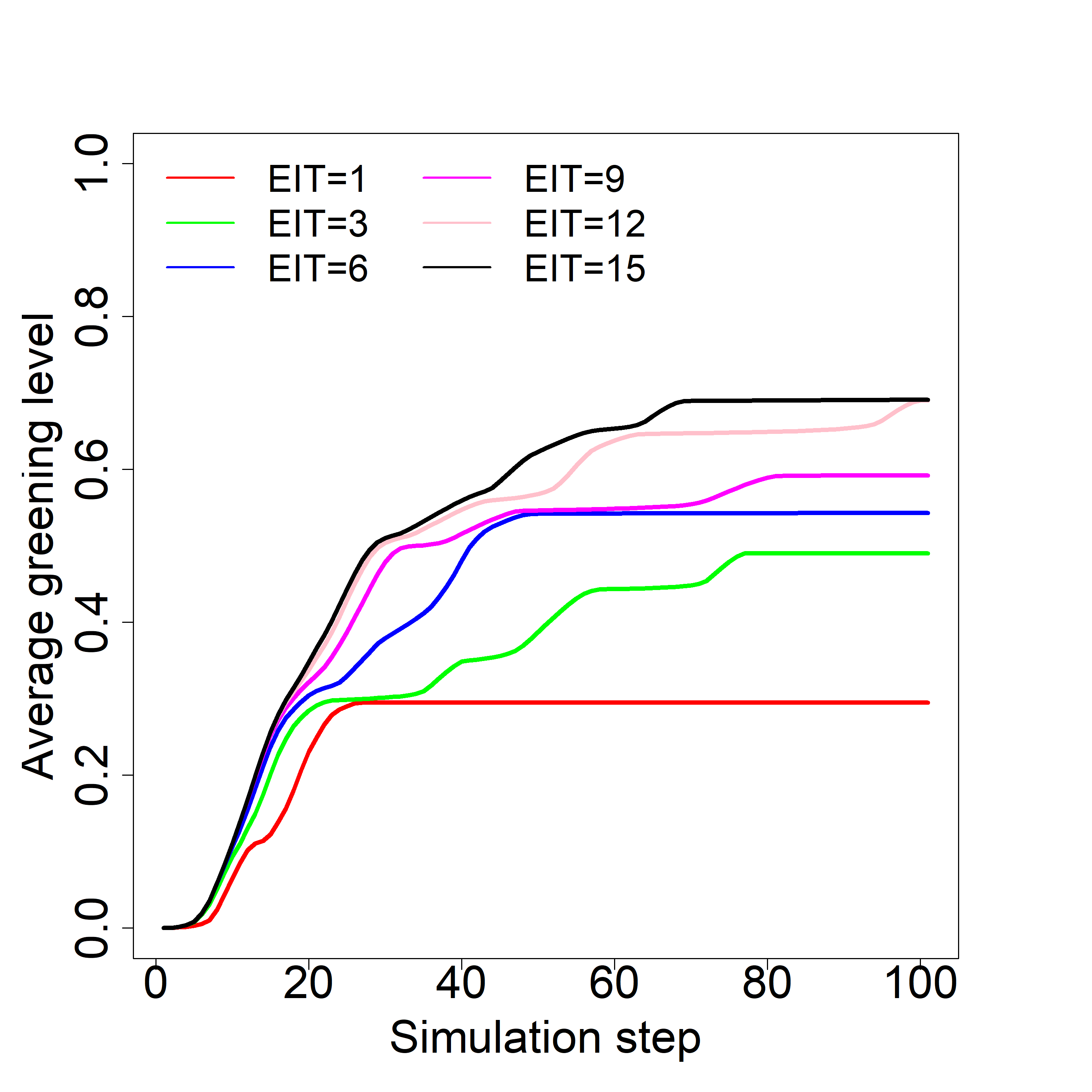}
  \caption{EIP=0.5}

\end{subfigure}
\begin{subfigure}[b]{0.31\textwidth}
  \includegraphics[width=\textwidth]{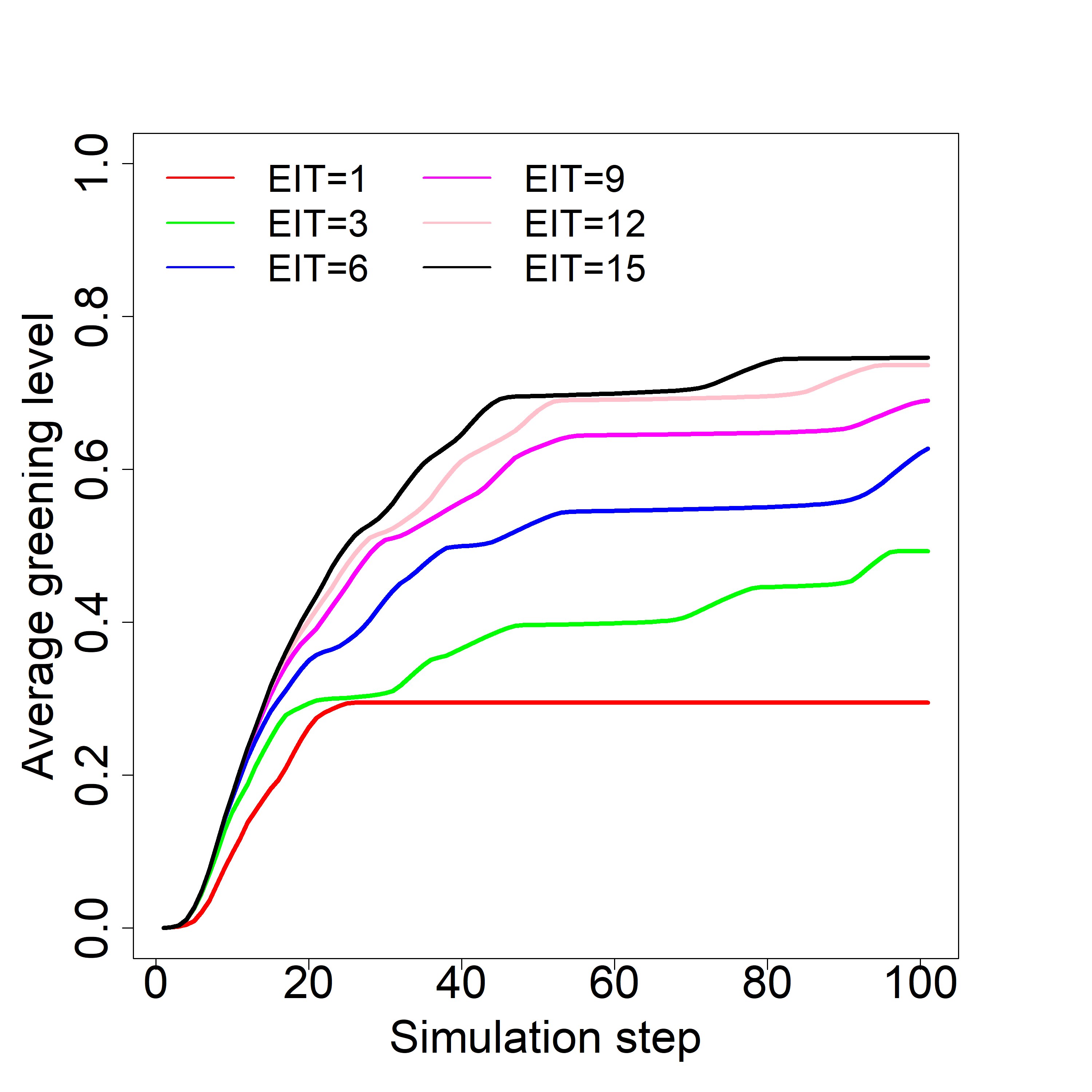}
  \caption{EIP=0.75}

\end{subfigure}
\begin{subfigure}[b]{0.31\textwidth}
  \includegraphics[width=\textwidth]{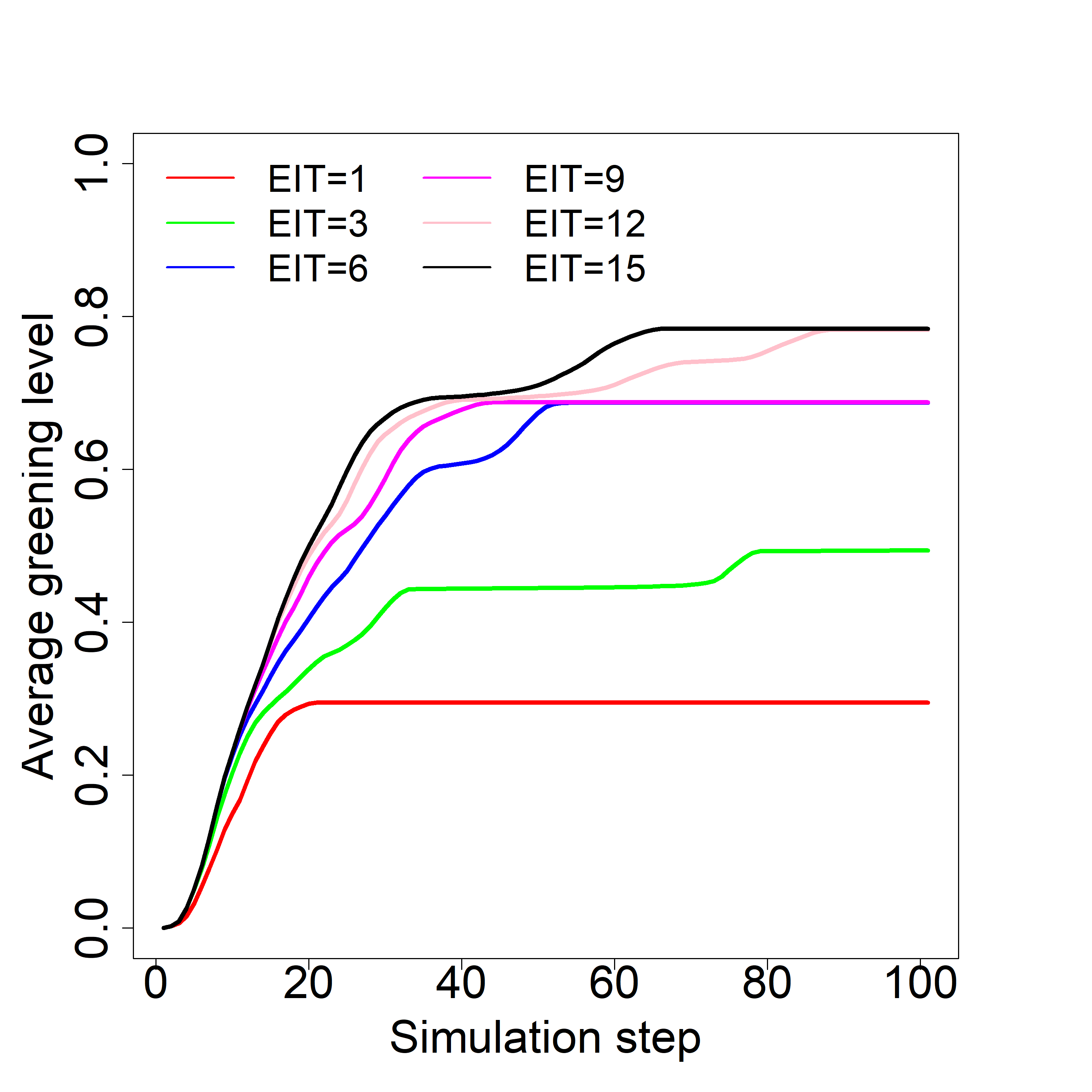}
  \caption{EIP=1.0}

\end{subfigure}
\begin{subfigure}[b]{0.31\textwidth}
  \includegraphics[width=\textwidth]{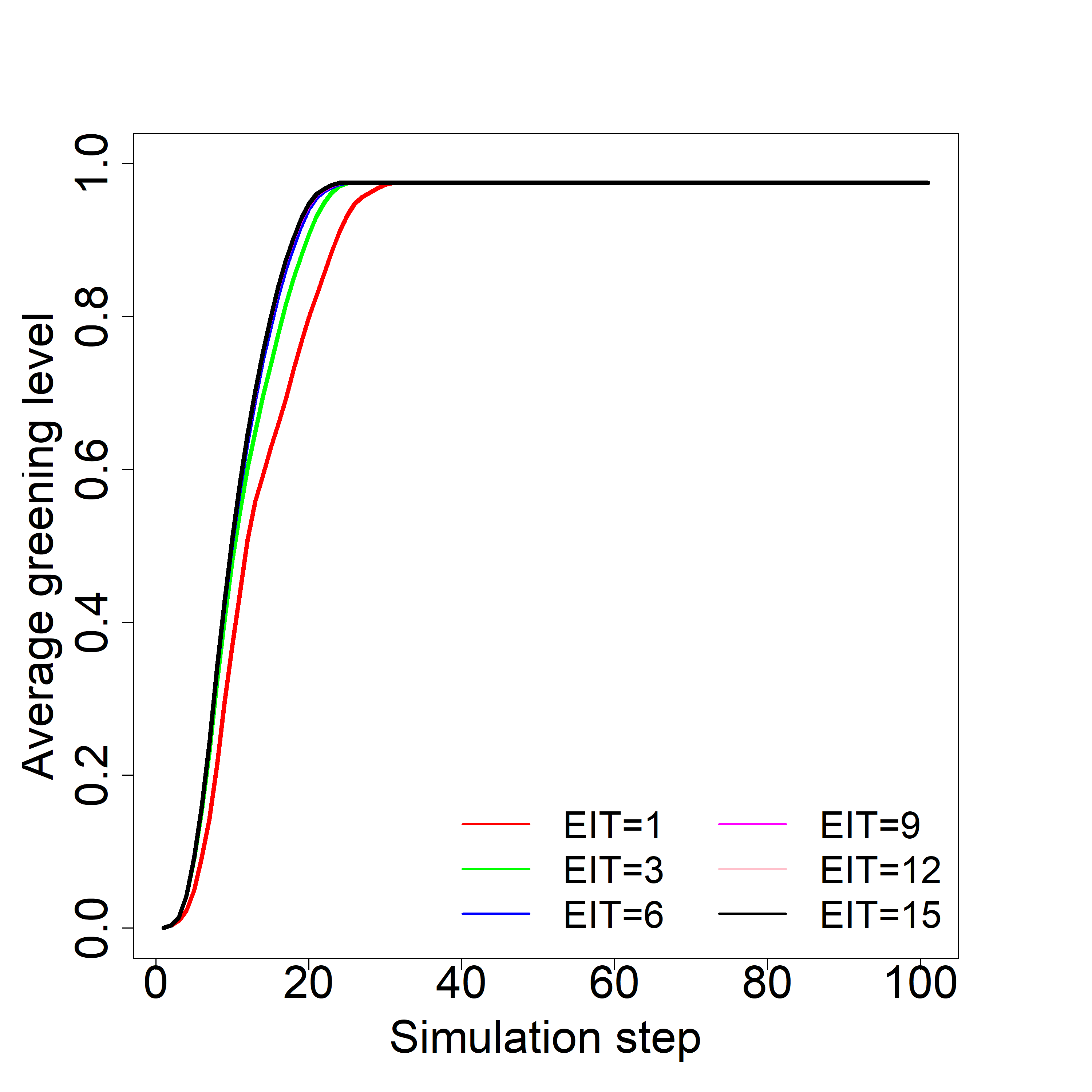}
  \caption{LT=0.05}

\end{subfigure}
\begin{subfigure}[b]{0.31\textwidth}
  \includegraphics[width=\textwidth]{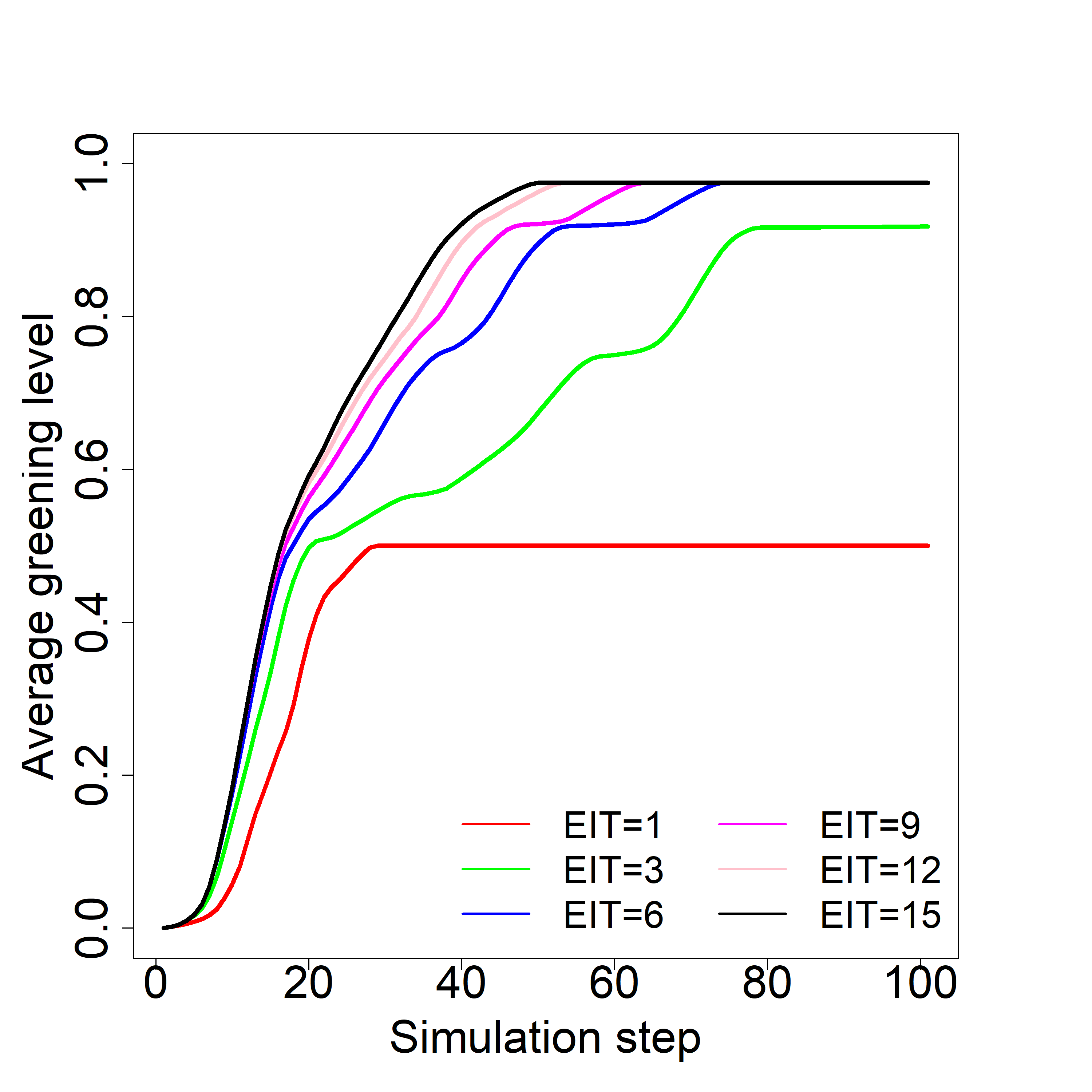}
  \caption{LT=0.1}

\end{subfigure}
\begin{subfigure}[b]{0.31\textwidth}
  \includegraphics[width=\textwidth]{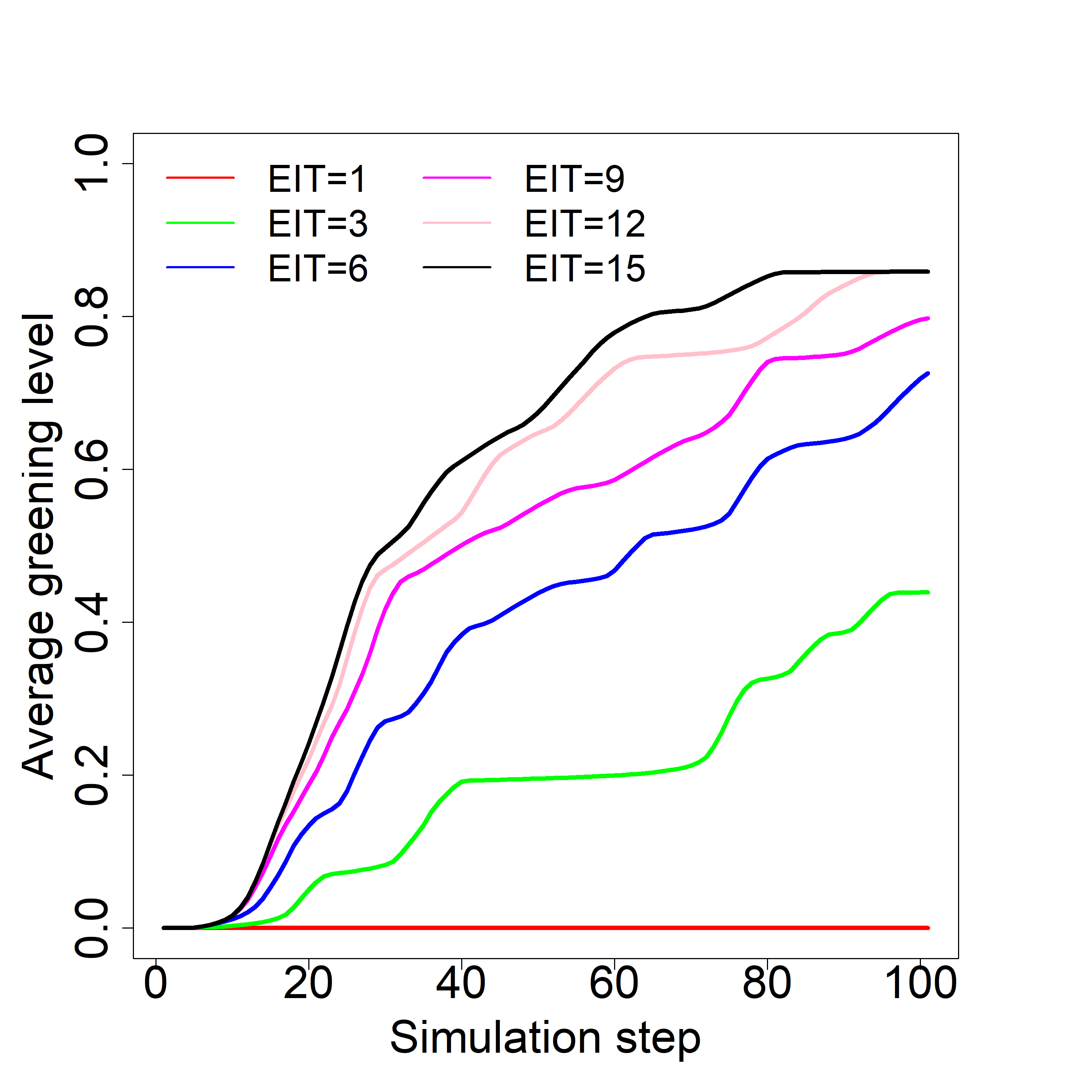}
  \caption{LT=0.15}

\end{subfigure}
\begin{subfigure}[b]{0.31\textwidth}
  \includegraphics[width=\textwidth]{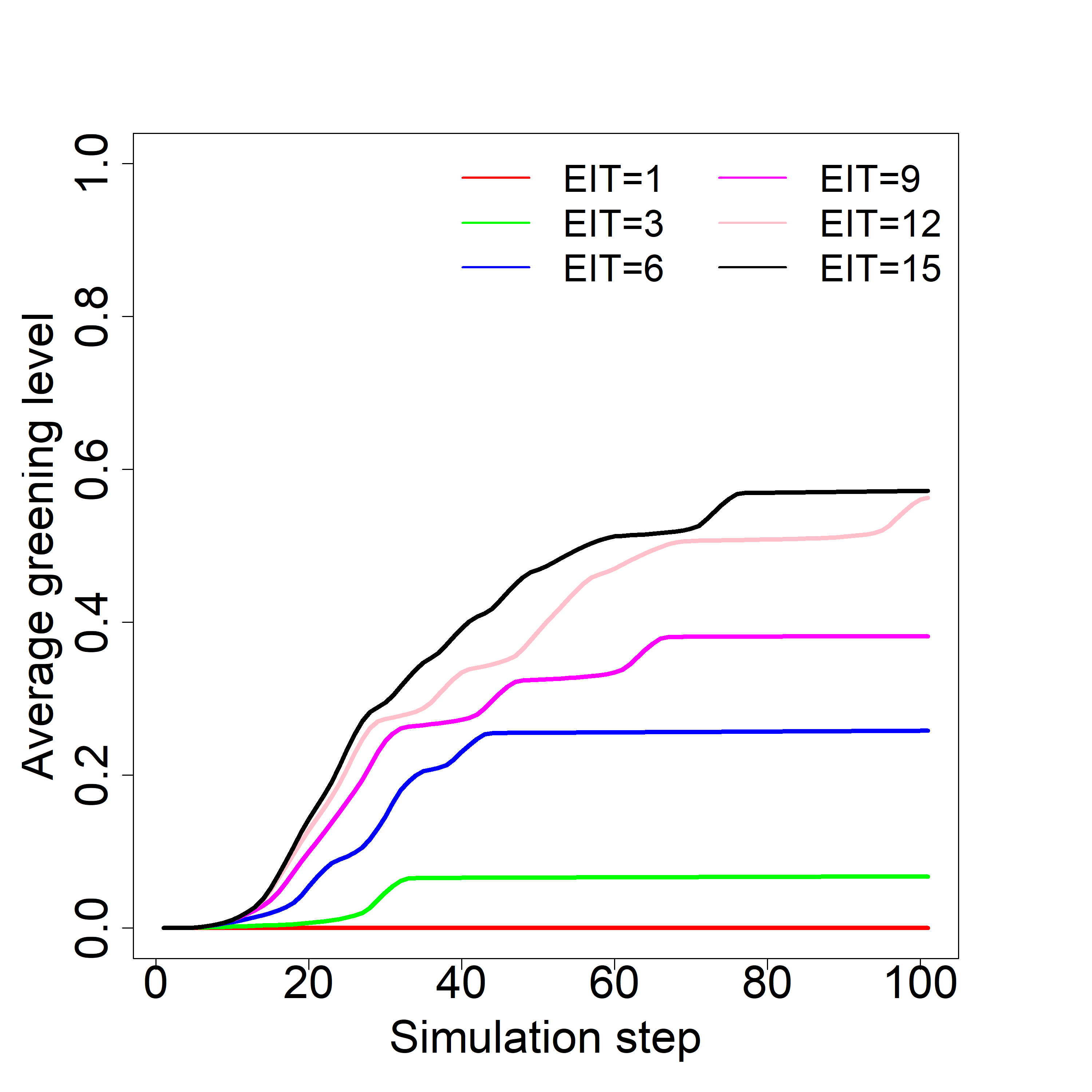}
  \caption{LT=0.2}

\end{subfigure}
\begin{subfigure}[b]{0.31\textwidth}
  \includegraphics[width=\textwidth]{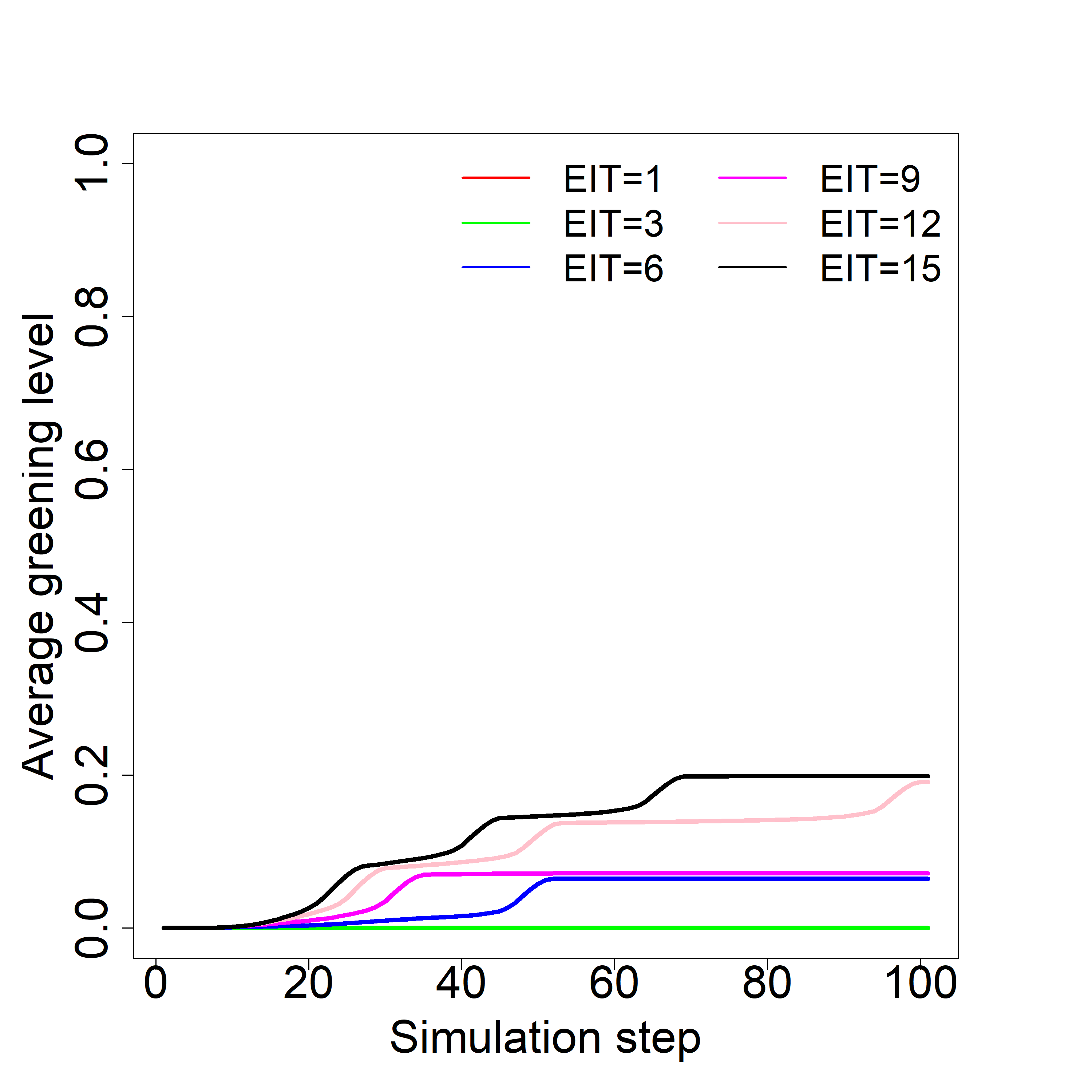}
  \caption{LT=0.25}

\end{subfigure}

\caption{Impact of External Influence Time $EIT$ on the greening level within companies for External Influence Probability ($EIP$)
(\textbf{a})~0.25, 
(\textbf{b})~0.50, 
(\textbf{c})~0.75, 
(\textbf{d})~1.00, and Linear Threshold ($LT$)
(\textbf{e})~0.05, 
(\textbf{f})~0.10, 
(\textbf{g})~0.15, 
(\textbf{h})~0.20, 
(\textbf{i})~0.25. 
}
\label{fig:eit_eip_lt}
\end{figure*}

In the next stage, the impact of $EIT$ was analysed for each used External Influence Probability $EIP$ with values 0.25, 050, 0.75 and 1.00. Figs.~\ref{fig:eit_eip_lt}(a-d) show the results for selected (for better clarity) six values of $EIT$ (1, 3, 6, 9, 12 and 15). $EIP$ influenced the average greening level and the dynamics of spreading processes for each $EIT$. For the lowest $EIP$~=~0.25 for most $EIT$ values, the processes reach saturation before the end of simulations. Higher the $EIP$ is, the saturation for most processes is reached faster for the same $EIT$ or the final average greening level is higher. It is especially visible for $EIT$~=~1, with saturation achieved around steps 20, and 30 for all used $EIP$ values. 

Analysis of impact of $LT$ on $EIT$ performance for each external influence time with is presented in figs.~\ref{fig:eit_eip_lt}(e-i). It shows that for the lowest $LT$~=~0.05 saturation at the level of average 0.98, green is achieved close to the 30th step of simulation for most of the $EIT$ values. This kind of social awareness requires little influence on banks to initiate processes, and no additional support is needed. 
Differences for $EIT$ are observed starting from $LT$~=~0.1. $EIT$~=~1 is reaching saturation at the level of 0.50 in the 27th step. $EIT$~=~3 allowed a much higher greening level of 0.92, and the process lasted longer to step 78. Processes with higher $EIT$ achieved a similar greening level of 0.98 but faster. Higher values of $LT$ resulted in a performance drop, saturation is reached for lower greening levels, and it happens faster, especially for $LT$~=~0.20 and $LT$~=~0.25.

The level of global support and demand for green technologies and the time of influence of external regulations have the greatest impact on the greening process. Therefore, it is important to build general acceptance for greening. Research on the German capital market shows that the main incentives for adopting sustainable finance include policy and legislative framework, investors' requirements and public pressure~\cite{gsfcg2018report}. There are many elements helpful in the creation of a change-friendly ecosystem, such as a stable institutional environment, systemic market education (including available platforms for the exchange of experience and knowledge), competitive financing (the introduction of new financial instruments supporting the implementation of pro-ecological projects and the transformation of enterprises towards sustainability), and expert support~\cite{deloitte2019perspektywy}. The time of impact of external regulations is also of significant importance for the greening process. Therefore, the introduced regulations should be well-thought-out, long-term and coordinated with the state's policy, especially in the field of the environment.

\section{Summary}
One of the greatest global challenges of recent years has been the search for effective ways of transitioning to sustainable development. The role of banks in greening the economy becomes extremely important. Banks can direct the flow of capital to the realization of pro-ecological projects and stimulate enterprises to green changes (e.g., costs of loans). The decision-makers and regulators have noticed the role of banks in the green transformation, which is why more regulations encourage or oblige banks to green behaviour.

In this study, we have evaluated how to influence green behaviour spreading processes within a multilayer network of banks and companies by external influence represented by market regulators. First, the new financial multilayer network model was proposed, and the green behaviour spreading model for financial multilayer networks was proposed. Finally, the green behaviour spreading model was used to assess the external influence on green behaviour spreading processes in the financial market. 

Results showed that external influence could be relatively low if a high level of education and positive global attitudes is maintained within the network. It was evident to analyze the impact of external influence time for different threshold values. Apart from the direct impact on financial institutions, if governments support social campaigns and educational aspects, efforts can be lower with a direct impact on the financial sector because of natural consumers' preferences. If global conditions don't support green behaviours, even long-term influence is not enough to achieve the highest greening level among companies which was visible, especially for higher thresholds. For low social and commercial support for the greening process, external influence may be one of the key options to improving green technology adoption. Lack of it stops adoptions and greening.  

Regulators may use lower impact on financial institutions with high global support and demand. Voluntary rules can be enough to achieve a high greening level. An increase in the strength of influence towards obligatory rules will not yield further improvements while high levels are achieved even without them. Increasing the strength of the impact of regulations delivers a growing increase together with increased resistance represented by higher thresholds. 

Increasing willingness to adopt green behaviour at the node level improved performance for all thresholds. Still, assuming the costs of increasing it for the lowest thresholds is not justified. Results show that increasing the willingness to adopt twice was more effective than increasing twice the probability of adoption based on the type of regulations. 

The selection of the target group is important for planning external impact on financial institutions. An increase in coverage is related to the growing costs of the campaign. Depending on global thresholds and network structures, addressing influence to a selected group of nodes can bring similar results based on higher costs targeting all of them. 

The results for used networks and set of parameters show that it is possible to achieve a high greening level for companies when the influence of regulators is focused on banks. The effect can be strengthened with external influence on both market participants. Strong demand from companies for green financial products creates demand for new offers. Obtained results require further investigation, especially regarding real banking systems and relations between companies and banks offering green products. 

In simulations, we have used the same $\alpha$, $\delta$ and $LT$ for all nodes. In real life, these values will differ depending on the type, size and other characteristics of a bank/company and other things like, for example, country. Thus, in our future work, we would like to model the distribution of $\alpha$, $\delta$ and $LT$ based on real data and study the impact of external influence on the market in a more realistic scenario.

\bibliographystyle{IEEEtran}
\bibliography{references}

\end{document}